\documentclass{article}
\pdfoutput=1
\usepackage{amsmath,amssymb,epsfig,color}
\numberwithin{equation}{section}
\usepackage{hyperref}

\DeclareMathAlphabet{\mathpzc}{OT1}{pzc}{m}{it}

\title{Double logarithms in ${\cal N} \geq 4$ supergravity:  \\
weak gravity and Shapiro's time delay } 
\author{Agust{\' \i}n Sabio Vera\\ 
\\
CERN, Theoretical Physics Department, Geneva, Switzerland.\\
\\
Instituto de F{\' \i}sica Te{\' o}rica UAM/CSIC, Nicol{\'a}s Cabrera 15\\ 
\& U. Aut{\' o}noma de Madrid, E-28049 Madrid, Spain.} 

\begin{document} 

\maketitle 

\begin{abstract}
A study of the double logarithmic in the center--of--mass energy, $s$, contributions to the four--graviton scattering amplitude is presented for four--dimensional ${\cal N} \geq 4$ supergravities. This includes a novel representation for the coefficients of the perturbative expansion based on exactly solvable recurrences. A review is given of the structure in the complex angular momentum plane for the $t$-channel partial wave singularities of the different amplitudes. 
Working in impact parameter representation, $\rho$, it is shown that the resummation of double logarithms makes gravity weaker in regions of small $\rho$ and large $s$. This screening of the gravitational interaction at short distances in the double logarithmic sector of the amplitudes is more acute as the number of gravitinos in the theory increases. It brings  corrections to the eikonal phase which can change the sign of the graviton's deflection angle and generate 
regions with repulsive interaction.  For very small impact parameters there appears a constant negative shift in both the eikonal phase and Shapiro's time delay which is not large enough  to generate causality violation. 
\end{abstract}

\section{Introduction}

The AdS/CFT correspondence~\cite{Maldacena:1997re,Gubser:1998bc,Witten:1998qj} and the physics program at the Large Hadron Collider (LHC) have revamped the interest in the calculation of scattering amplitudes. Being fundamental for the former and an interesting simplified model for the latter, ${\cal N} = 4$ supersymmetric Yang-Mills theory has played a pivotal role in recent developments. A very interesting finding~\cite{Bern:2002kj} has been the double-copy structure linking this theory to ${\cal N} = 8$ supergravity~\cite{Bern:1998ug,Bern:2009kd}.

An interesting limit where to investigate scattering amplitudes in gravitational theories, of relevance in the work here presented, is that of multi-Regge kinematics. In this case the center--of--mass energy squared, $s$, is asymptotically larger than all other Mandelstam invariants and amplitudes factorize in terms of impact factors and reggeized gravitons exchanged in the $t$-channel~\cite{Grisaru:1975tb,Grisaru:1981ra}. For gravity,  eikonal contributions are also relevant together with double-logarithmic in $s$ (DL) terms~\cite{Lipatov:1982vv,Lipatov:1982it,Lipatov:1991nf}. Both can be studied using the high energy effective action derived by Lipatov~\cite{Lipatov:2011ab}. In this framework the leading contributions to scattering amplitudes are generated by bunches of produced gravitons  well separated in rapidity from each other. The regions in rapidity without radiation have their origin in the exchange of reggeized gravitons. These are created (annihilated) in the $t$-channel by $A^{++} (A^{--})$ fields, which are subject to the constraints $\partial_\pm A^{\pm \pm} =0 $ in order to generate the rapidity gaps. The local and non-local (in rapidity) effective interactions which appear in this context are described by the action given in~\cite{Lipatov:2011ab}. It contains the Einstein-Hilbert action together with a kinetic term for the reggeon fields plus induced contributions.  Within this setup it is possible to calculate the graviton Regge trajectory and those effective vertices needed to produce inelastic amplitudes. 
 
Connecting with the double--copy structure of gravity~\cite{Vera:2014tda},  it is remarkable that the graviton emission vertex in multi--Regge kinematics can be written as a modified double copy of the equivalent~\cite{BFKL1,BFKL2,BFKL3} gluon emission vertex in QCD~\cite{Lipatov:1982vv,Lipatov:1982it,Lipatov:1991nf,SabioVera:2011wy}.

The interplay between the reggeization of the graviton and the DL contributions (DLs) is not trivial. The DLs in Einstein--Hilbert gravity and supergravity, which will be further studied in the present work, were originally calculated in~\cite{Bartels:2012ra}.  Those results are in agreement up to two loops for ${\cal N}=4,5,6,8$ supergravities with the double--copy study presented in~\cite{BoucherVeronneau:2011qv}.  More recently, further confirmation of~\cite{Bartels:2012ra} has been found at three loops for ${\cal N} = 8$   supergravity in~\cite{Henn:2019rgj}. The DL amplitudes in this theory have been investigated in impact parameter representation, together with their mapping to the problem of counting 1--rooted ribbon graphs, in~\cite{SabioVera:2019edr}. 

If $L$ is the loop order, $N$ the number of gravitinos of the theory, $s=(p_1+p_2)^2$, $t=(p_1-p_3)^2$, $u=(p_1-p_4)^2$ and $\kappa^2 = 8 \pi G = 8 \pi^2 \alpha$, with $G$ being Newton's constant, the 
 four--graviton scattering amplitude with helicities (++;++) can be written in the form 
\begin{eqnarray}
{\cal A}_{4,(N)} &=& {\cal A}_4^{\rm Born} {\cal M}_{4,(N)}
~=~ \kappa^2 \frac{s^3}{t u} \left(1 + \sum_{L=1}^\infty {\cal M}_{4,(N)}^{(L)}\right).
\label{eqnnotation}
\end{eqnarray}
In the Regge limit, $s \gg -t={\vec{q}}^{\, 2}$, Lipatov found~\cite{Lipatov:1982vv,Lipatov:1982it,Lipatov:1991nf} that the one--loop amplitude contains the graviton Regge trajectory, 
\begin{eqnarray}
\omega (q) &=& \frac{\alpha \, {\vec{q}}^{\, 2}}{\pi}\int \frac{d^2 \vec{k} }{\vec{k}^2 
(\vec{q}-\vec{k})^2}\nonumber\\
&\times&  \left((\vec{k} \cdot (\vec{q}-\vec{k}))^2\left(\frac{1}{\vec{k}^2 }+
\frac{1}{(\vec{q}-\vec{k})^2}\right)+\frac{N}{2}\,(\vec{k} \cdot (\vec{q}-\vec{k}))-{\vec{q}}^{\, 2}\right).
\label{onelooptraj}
\end{eqnarray}
Its infrared divergence can be regularized by a cut--off $\lambda$ and the ultraviolet one by $s$. The dependence on $\lambda$ is well known in gauge theories and should cancel when evaluating physical observables. The ultraviolet divergence is due to the use of an effective high energy theory and has a kinematical origin. Eq.~(\ref{onelooptraj}) then becomes
\begin{eqnarray}
\omega (q) &=& \alpha \,t 
\,\left(\ln \left(\frac{-t}{\lambda ^2}\right) +\frac{N-4}{4}\,\ln \left(\frac{s}{-t}\right)\right)\,.
\label{trajgrav}
\end{eqnarray}
This result should be compared to the exact one--loop amplitudes. The simplest case is ${\cal N}=8$ supergravity, where $N=8$, for which
\begin{eqnarray}
{\cal M}^{(1)}_{4, (N=8)}  &=& \alpha \, t \ln{\left(\frac{-s}{-t}\right)}\ln{\left(\frac{-u}{-t}\right)} \nonumber\\
&+&\alpha \, \frac{t}{2} \ln{\left(\frac{-t}{\lambda^2}\right)}
\left(\ln{\left(\frac{-s}{-t}\right)}+\ln{\left(\frac{-u}{-t}\right)}\right)  \nonumber\\
&-& \alpha \frac{(s-u)}{2} \ln{\left(\frac{-t}{\lambda^2}\right)}\ln{\left(\frac{-s}{-u}\right)} \, .
\label{oneloopn8exact}
\end{eqnarray}
Two--loop corrections to this formula for different supergravities can be found in~\cite{BoucherVeronneau:2011qv} and at one--loop for usual gravity in~\cite{Dunbar:1994bn}. They are in agreement with Eq.~(\ref{trajgrav}). 
At high energies $u \simeq -s$ and the first line in Eq.~(\ref{oneloopn8exact}) 
contains a double logarithm in energy of the form $\alpha \,  t \ln^2{s}$. The analysis of the resummation of these terms for ${\cal N} \geq 4$ supergravities is the target of the work here presented.

Lipatov calculated~\cite{Lipatov:1982vv} the DLs present to all orders in the four-graviton amplitude generated by ladder diagrams. In~\cite{Bartels:2012ra} it was shown that non-ladder contributions are important and should be included in the calculation. Non-ladder terms appeared before in quantum electrodynamics (QED).  For example, in $e^+e^-$ forward scattering they take the form of a Bessel function. In the case of $e^+e^-$ backward scattering Sudakov photon lines should be attached to the external fermions generating an equation with a parabolic cylinder function as solution. These results were obtained by Gorshkov, Gribov, Lipatov and Frolov long ago~\cite{Gorshkov:1966ht,Gorshkov:1966qd,Gorshkov:1966hu}. Similar techniques can be used in quantum chromodynamics (QCD), applied to deep inelastic scattering~\cite{Kirschner:1982qf,Kirschner:1982xw,Kirschner:1983di}, or in the electroweak sector of the Standard Model~\cite{Fadin:1999bq}.

In the next section a novel representation of the DL terms for the four--graviton scattering amplitude in different supergravities is given and studied in detail. 
After this the associated singularity structure in the complex angular momentum plane for the partial waves 
is explained, together with a numerical study of the position of their poles. Finally, the representation of the amplitudes in impact parameter space is introduced highlighting 
the role played by the DL corrections in the eikonal phase, having as a consequence a shift in the  
Shapiro's time delay and the change of sign in the graviton's deflection angle in the forward limit. Some conclusions are drawn at the end. 

\section{Double logarithms in perturbation theory}

To work with the four--graviton scattering amplitude in four--dimensional supergravity, at  DL accuracy, it is convenient to extract the $\lambda$--dependent piece in the graviton Regge trajectory and to use the representation
\begin{eqnarray}
{\cal A}_{4,(N)}  &=& {\cal A}_4^{\rm Born}\,\left(\frac{s}{-t}\right)^{\alpha t\,\ln \left(\frac{-t}{\lambda ^2}\right)}
\int _{\delta-i\infty}^{\delta +i\infty}\frac{d\,\omega }{2\pi i}\,
\left(\frac{s}{- t}\right)^\omega \frac{ f ^{(N)}_\omega}{\omega} \,,\,\,\delta>0\, .
\label{factor}
\end{eqnarray}
The focus in the present work is the detailed study of the DLs in 
\begin{eqnarray}
{\cal M}_{4,{\rm DL}}^{(N)}   (s,t) &=& \int _{\delta-i\infty}^{\delta +i\infty}\frac{d\,\omega }{2\pi i}\,\left(\frac{s}{-t}\right)^\omega \frac{ f _\omega^{(N)} }{\omega} \, ,
\label{M4fomega}
\end{eqnarray}
where the perturbative expansion of the $t$-channel partial wave $f ^{(N)}_\omega$ reads
\begin{eqnarray}
f^{(N)}_\omega &=& \sum _{n=0}^\infty {\cal C}_n^{(N)}\,\left(\frac{\alpha t}{\omega^2}\right)^n\,.
\label{fomegaexp}
\end{eqnarray}
In terms of Eq.~(\ref{M4fomega}) this implies
 \begin{eqnarray}
{\cal M}_{4,{\rm DL}}^{(N)}  (s,t) &=&  1+ \sum_{n=1}^\infty {\cal C}_n^{(N)} \, 
\frac{(\alpha \, t)^n}{(2n)!} \ln^{2n}{\left(\frac{s}{-t}\right)} \, .
\label{PertAmplDL4}
\end{eqnarray} 

In~\cite{Bartels:2012ra} it was shown that the ladder and non-ladder DLs to the four--graviton amplitude are generated by the solution to the 
equation, similar to the QED and QCD cases, for the partial wave
\begin{eqnarray}
f_\omega^{(N)}  &=& 1+ \alpha\,  t \left(\eta_N \frac{f_\omega^{(N) 2}}{\omega ^2}-\frac{d}{d\,\omega}\,\frac{f_\omega^{(N)} }{\omega}\right)
\label{fomegaTheEq}
\end{eqnarray}
where $\eta_N \equiv {N -6 \over 2}$. The derivative stems from virtual gravitons with the smallest $p_T$ and the quadratic term from low energy gravitons and gravitinos exchanged in the $t$-channel. Introducing the expansion (\ref{fomegaexp}) in Eq.~(\ref{fomegaTheEq}) it is possible to show that the coefficients follow the recursive relation,
\begin{eqnarray}
 {\cal C}_n^{(N)} &=&  (\eta_N+2n-1) \, {\cal C}_{n-1}^{(N)}      +  \eta_N\sum _{j=1}^{n-1} {\cal C}_j^{(N)} {\cal C}_{n-1-j}^{(N)}  , \, \, \, {\cal C}_0^{(N)} = 1 \, . 
 \label{recursion}
\end{eqnarray}
The sequence of terms can be obtained by iteration, {\it e.g.}
\begin{eqnarray}
 {\cal C}_0^{(N)} &=&1 \, ,\nonumber\\
  {\cal C}_1^{(N)} &=&\eta_N +1\nonumber\\
 &=&\frac{N-4}{2} \, ,\nonumber\\
  {\cal C}_2^{(N)} &=&  2 \eta_N^2+5 \eta_N +3\nonumber\\
 &=&\frac{(N-4)}{2}  (N-3)\, ,\nonumber\\
  {\cal C}_3^{(N)} &=&5 \eta_N ^3+22 \eta_N ^2+32 \eta_N +15\nonumber\\
 &=&\frac{(N-4)}{8}  \left(5 N^2-26 N+36\right)\, ,   \label{Sixterms}\\
  {\cal C}_4^{(N)} &=&14 \eta_N ^4+93 \eta_N ^3+234 \eta_N ^2+260 \eta_N +105 \nonumber\\
  &=& \frac{(N-4)}{8}  \left(7 N^3-47 N^2+118 N-108\right)\, ,\nonumber\\
  {\cal C}_5^{(N)} &=&42 \eta_N ^5+386 \eta_N ^4+1450 \eta_N ^3+2750 \eta_N ^2+2589 \eta_N +945\nonumber\\
 &=& \frac{(N-4) }{16} \left(21 N^4-160 N^3+556 N^2-960 N+648\right)\, ,\nonumber\\
  {\cal C}_6^{(N)} &=&132 \eta_N ^6+1586 \eta_N ^5+8178 \eta_N ^4+22950 \eta_N ^3+36500 \eta_N ^2+30669
   \eta_N +10395 \nonumber\\
   &=& \frac{(N-4)}{16}  \left(33 N^5-263 N^4+1156 N^3-2828 N^2+3576
   N-1944\right) \, .\nonumber
\end{eqnarray}

The recursive relation for the $N=4$ coefficients has a trivial solution where only the first one survives and is equal to one. This shows that the DLs are not present in ${\cal N}=4$ supergravity. There must be a symmetry reason for this cancellation and work is in progress to understand this fact.  As discussed in~\cite{Bartels:2012ra},  ${\cal N}=4$ supergravity is a critical theory which marks the transition from badly behaved amplitudes at high energies when $N<4$, which blow up at large $s$, to finite amplitudes which manifest a convergent  behaviour when $N>4$. For this second class of theories it is possible to find closed expressions for the coefficients of the DL sector in the four--graviton amplitude.  

As explained in~\cite{SabioVera:2019edr}, there is a rich history for the coefficients ${\cal C}_n^{(8)}$. They are present in many body theory~\cite{Ihrig:1976ih}, as combinatoric weights in QED~\cite{Cvitanovic:1978wc} and in calculations of the polaron Green's function in solid state physics~\cite{Goodvin:2006Go}. The non-linear equation~(\ref{recursion}) belongs to a class of exactly solvable recurrences studied in mathematical works on combinatorics. In~\cite{MartinKearney:2011} the self-convolutive recurrence relations 
\begin{eqnarray}
u_n &=& (\alpha_1 n + \alpha_2) u_{n-1} + \alpha_3 \sum_{j=1}^{n-1} u_j u_{n-j}, \,\, \, u_1 ~=~ 1 \, ,
\label{generalueqn}
\end{eqnarray}
were solved in the form $u_n = \int_0^\infty x^{n-1} \mu (x) dx$. Provided that $(\alpha_1,\alpha_2,\alpha_3) = (2,-3,\eta_N)$ and  $u_{n+1} = {\cal C}_n^{(N)}$, it follows that Eq.~(\ref{recursion}) is equivalent to Eq.~(\ref{generalueqn}). Using the representation 
\begin{eqnarray}
{\cal C}_n^{(N)} &=& \int_0^\infty x^{n} \mu^{(N)} (x) \, dx
\label{coeffsCnN}
\end{eqnarray}
it is then possible to write 
\begin{eqnarray}
\mu^{(N)} (x) &=& \frac{ \frac{2^{1-\eta_N}}{\eta_N } 
\Gamma   \left(\eta _N\right)
     \sqrt{\frac{2 e^{x }}{\pi x }} }{\Gamma \left(\frac{\eta
   _N}{2}\right){}^2 \, _1F_1\left(\frac{1-\eta
   _N}{2};\frac{1}{2};\frac{x}{2}\right){}^2+2 \, x \, \Gamma \left(\frac{1+\eta _N}{2}
   \right){}^2 \, _1F_1\left(\frac{2-\eta
   _N}{2};\frac{3}{2};\frac{x}{2}\right){}^2}  
   \label{muNncoeffs} 
\end{eqnarray}
where $\Gamma$ is the usual gamma function and $_1F_1$ is the Kummer confluent hypergeometric function, defined as the solution to
\begin{eqnarray}
z \, {_1F_1}^{\prime \prime}
+\left(b-z\right) \, {_1F_1}^{\prime}-a \, _1F_1 ~=~ 0 \, . 
\end{eqnarray}
It can be written in terms of the expansion
\begin{eqnarray}
_1F_1(a;b;z) ~=~  \sum_{n=0}^\infty \frac{(a)_n}{(b)_n} \frac{z^n}{n!}\, , \,\, 
 (a)_n ~=~ \frac{\Gamma(a+n)}{\Gamma(a)} \, ,
   \end{eqnarray}
with $(a)_n$ being the Pochhammer symbol.   

This is a novel closed representation of the perturbative coefficients for different supergravities which allows to write the resummation of the DLs to the scattering amplitude in the form
 \begin{eqnarray}
{\cal M}_{4,{\rm DL}}^{(N)}  (s,t) &=&   
 \int_0^\infty \, dx \, \mu^{(N)} (x) 
\cosh \left(  \sqrt{\alpha \, t \, x} \ln \left(\frac{s}{-t}\right)\right) \, .
\label{M4DLcosh}
\end{eqnarray} 
The weight function in Eq.~(\ref{muNncoeffs}) for each supergravity with 
$N>4$ becomes
\begin{eqnarray}
\mu^{(5)} (x) &=& \frac{16 \sqrt{\frac{e^x}{x}}}{\Gamma\left(-\frac{1}{4}\right)^2 \, _1F_1\left(\frac{3}{4};\frac{1}{2};\frac{x}{2}\right){}^2+2 \, x \, \Gamma \left(\frac{1}{4}\right)^2 \, _1F_1\left(\frac{5}{4};\frac{3}{2};\frac{x}{2}\right){}^2} \, ,
\label{muN5}\\
\mu^{(6)} (x) &=& \frac{1}{ \sqrt{2 \pi x e^x }} \, ,
\label{muN6}\\
\mu^{(7)} (x) &=& \frac{\frac{4 \sqrt{2} }{\pi ^2 x} }{
   I_{-\frac{1}{4}}\left(\frac{x}{4}\right){}^2+I_{
   \frac{1}{4}}\left(\frac{x}{4}\right){}^2}\, ,
   \label{muN7}\\
   \mu^{(8)} (x) &=&  \frac{ \sqrt{\frac{2 e^x}{\pi^3 x}}}{\text{erfi}\left(\sqrt{\frac{x}{2}}\right
   )^2+1}\, ,
   \label{muN8}
\end{eqnarray}
where $I_{a}(z)$ is the modified Bessel function of the first kind and ${\rm erfi} (z)$ the imaginary error function. 
These expressions generate the set of coefficients ${\cal C}^{(N)} = 
\left({\cal C}_0^{(N)},{\cal C}_1^{(N)}, \dots \right)$ for each supergravity theory in agreement with Eq.~(\ref{Sixterms}). More explicitly,
\begin{eqnarray}
{\cal C}^{(4)} &=&  \left(1,0,0,0,0,0,0, \dots\right) \, ,\\
{\cal C}^{(5)} &=&  \left(1,\frac{1}{2},1,\frac{31}{8},\frac{91}{4},\frac{2873}{16},\frac{14243}{8}, \dots\right) \, ,\\
{\cal C}^{(6)} &=&  \left(1,1,3,15,105,945,10395, \dots\right) \, ,\\
{\cal C}^{(7)} &=&  \left(1,\frac{3}{2},6,\frac{297}{8},306,\frac{50139}{16},38286, \dots\right) \, , \\
{\cal C}^{(8)} &=&  \left(1,2,10,74,706,8162,110410 \dots\right) \, .
\end{eqnarray}

As explained above, for ${\cal N}=4$ supergravity the double logarithms are not present. The only non-zero coefficient is that corresponding to the Born amplitude, ${\cal C}_n^{(4)} =  \delta_n^0$ and the scattering amplitude is constant in the DL approximation,  ${\cal M}_{4,{\rm DL}}^{(4)}  (s,t)  = 1$. For $N=6$ it is possible to perform the integration in Eq.~(\ref{coeffsCnN}) to obtain
\begin{eqnarray}
{\cal C}_n^{(6)} &=&   \frac{2^{n}}{\sqrt{\pi}}  
\int_0^\infty  e^{-x} x^{n-\frac{1}{2}} dx 
~=~ \frac{2^{n}}{\sqrt{\pi}} \, \Gamma \left(n+\frac{1}{2}\right) \, . 
\end{eqnarray}
The corresponding DL amplitude can therefore be calculated explicitly,
 \begin{eqnarray}
{\cal M}_{4,{\rm DL}}^{(6)}  (s,t)
&=& \int_0^\infty  dx \,   \frac{e^{-x} }{\sqrt{\pi \, x}}
 \cosh{    \left(   \sqrt{   2 \, \alpha \, t \,  x    } \ln {\left(\frac{s}{-t}\right)   }\right)}  ~=~  e^{\frac{\alpha}{2} t \ln^{2}{\left(\frac{s}{-t}\right)}},
\end{eqnarray}
generating a direct exponentiation of the DLs. 

\begin{figure}
\begin{center}
\includegraphics[width=9cm]{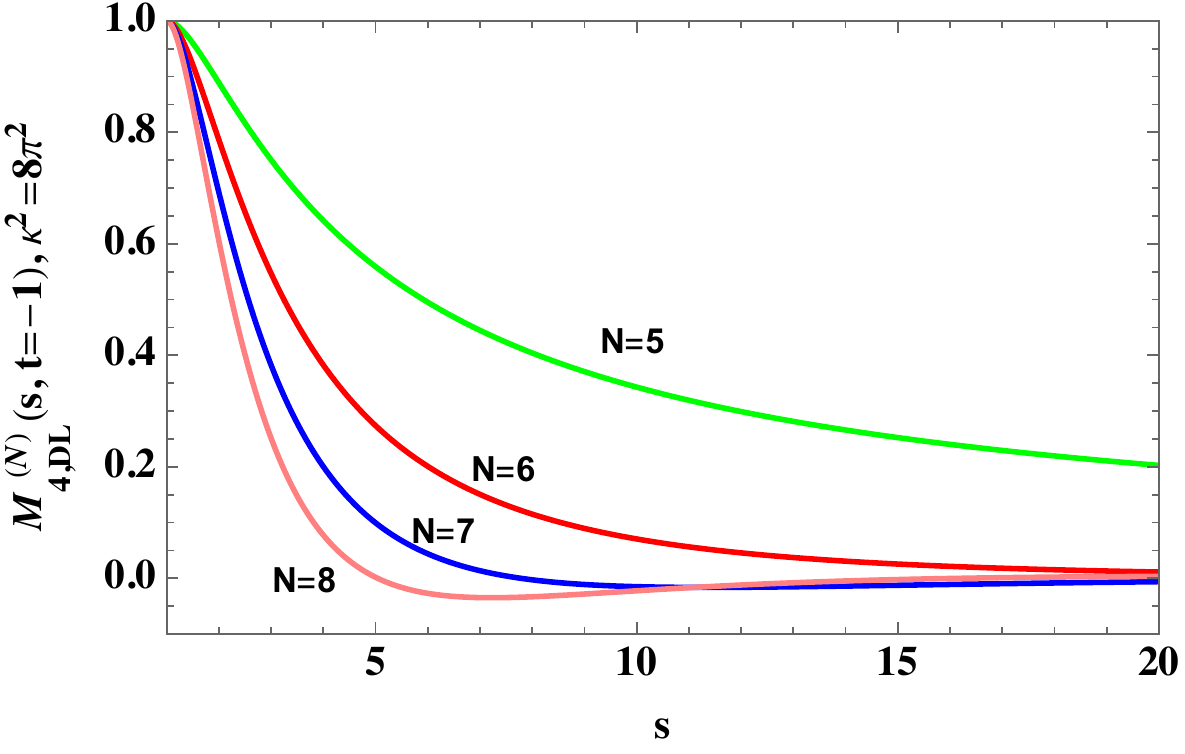}
\vspace{-.5cm}
\end{center}
\caption{The DL exact amplitudes for different supergravities with $N=5,6,7,8$  as a function of $s$ with $\kappa^2 = 8 \pi^2$ and $t = -1$. }
\label{DLExactAmplitudes}
\end{figure}

A plot of the different DL amplitudes with $\kappa^2 = 8 \pi^2$ and $t = -1$ is shown 
in Fig.~\ref{DLExactAmplitudes}. The $N=7,8$ amplitudes tend to zero very rapidly as $s$ increases featuring some oscillatory behaviour. 
In the $N=5,6$ cases they also tend to zero but more slowly and without oscillations. In the next section 
this is explained in terms of the singularity structure of their $t$--channel partial waves. 
All these plots have been produced by the numerical integration of Eq.~(\ref{M4DLcosh}) and they are equivalent to those presented in~\cite{Bartels:2012ra}, which were obtained by numerical integration of the partial waves over a contour in the complex angular momentum plane.  

\section{Singularity structure}
 
Beyond a purely perturbative treatment of the solution, the partial wave in 
Eq.~(\ref{M4fomega}) can be written in the form
\begin{eqnarray}
\frac{f_\omega^{(N)}}{\omega} &=& \frac{-1}{\eta_N} \frac{d}{d \omega} 
\ln{\phi_\omega^{(N)}}
\end{eqnarray}
with 
\begin{eqnarray}
\phi_\omega^{(N)} &=& \int_0^\infty} dz \, z^{\frac{N-8}{2}} e^{- \frac{z^2}{2}} e^{- \frac{\omega}{\sqrt{- \alpha t}}z \nonumber\\
&=& 2^{\frac{\eta_N }{2}-1} \Gamma \left(\frac{\eta_N }{2}\right) \,
   _1F_1\left(\frac{\eta_N }{2};\frac{1}{2};\frac{-\omega ^2}{2 t \alpha
   }\right) \nonumber\\
&+& 2^{\frac{\eta_N }{2}-\frac{1}{2}}  \frac{\omega  }{\sqrt{- \alpha t}}
\Gamma   \left(\frac{1+\eta_N }{2}\right) \, _1F_1\left(\frac{\eta_N
   +1}{2};\frac{3}{2};\frac{-\omega ^2}{2 t \alpha }\right) \, .
\end{eqnarray}
An equivalent, although simpler, representation is
\begin{eqnarray}
\frac{f_\omega^{(N)}}{\omega } &=&   \frac{1}{3-\frac{N}{2}}
\frac{d}{d \omega} \ln{H_{3-\frac{N}{2}}\left(\frac{w}{\sqrt{-2 \alpha t}}\right)}~=~ \frac{\sqrt{2} H_{2-\frac{N}{2}}\left(\frac{\omega }{ \sqrt{- 2 \alpha t
   }}\right)}{\sqrt{- \alpha  t} H_{3-\frac{N}{2}}\left(\frac{\omega }{ \sqrt{-2   \alpha t}}\right)}
\label{SolutionFomega}
\end{eqnarray}
where $H_n$ are the Hermite polynomials. In terms of a contour integral over a counterclockwise directed path from $\infty+i \,0$ to $\infty-i \,0$, $L$, enclosing the origin, they can be written in the form
\begin{eqnarray}
H_n (z) &=& 2^{n \over 2} e^{z^2 \over 2} D_n \left(\sqrt{2} z\right) \, , \\
D_n (z) &=& (-1)^{n}\Gamma(n+1) e^{- z^2 \over 4} \int_L \frac{dt}{2 \pi i} 
\frac{e^{-zt-\frac{t^2}{2}}}{t^{n+1}} \, ,
\end{eqnarray}
where $D_n$ is the parabolic cylinder function. 

This representation allows for the analytic continuation away from the real axis in $n$ and it is compatible with the perturbative expansion of $f_\omega^{(N)}$. To see this it is needed to use the asymptotic expansion for the parabolic cylinder function at $|z|\to \infty$: 
\begin{eqnarray}
D_\nu (z) &\sim&  e^{- z^2 \over 4} 
\sum_{n=0}^\infty \frac{(-1)^n (- \nu)_{2n}}{n! \, 2^n z^{2n-\nu}} 
\end{eqnarray} 
which implies
\begin{eqnarray}
H_\nu \left(z\right) &\sim&     
\sum_{n=0}^\infty \frac{(-1)^n (- \nu)_{2n}}{n!  \left(2 z\right)^{2n-\nu}} \, .
\end{eqnarray}
When applied to Eq.~(\ref{SolutionFomega}) this leads to the expansion in 
Eq.~(\ref{fomegaexp}),
\begin{eqnarray}
 \frac{ \sqrt{2} \omega } {\sqrt{-  \alpha  t} }
\frac{H_{2-\frac{N}{2}}\left(\frac{\omega }{ \sqrt{- 2 \alpha t
   }}\right) }{H_{3-\frac{N}{2}}\left(\frac{\omega }{ \sqrt{-2   \alpha t}}\right)}
 &\sim&  \sum _{n=0}^\infty {\cal C}_n^{(N)} \,\left(\frac{ \alpha \, t}{\omega^2}\right)^n \, . 
\end{eqnarray}

The expression in Eq.~(\ref{SolutionFomega}) reveals the singularity structure of the partial wave in the $\omega$ plane. It corresponds to a set of simple poles situated at the zeroes of the function 
$H_{3-\frac{N}{2}}$, which are placed in the left hand side of the complex plane.  Due to Eq.~(\ref{M4fomega}), the  asymptotic behaviour of the DLs to the scattering amplitude in the $s \gg -t$ limit is governed by the poles closest to the origin. 

\begin{figure}[h]
\begin{center}
\includegraphics[width=9cm]{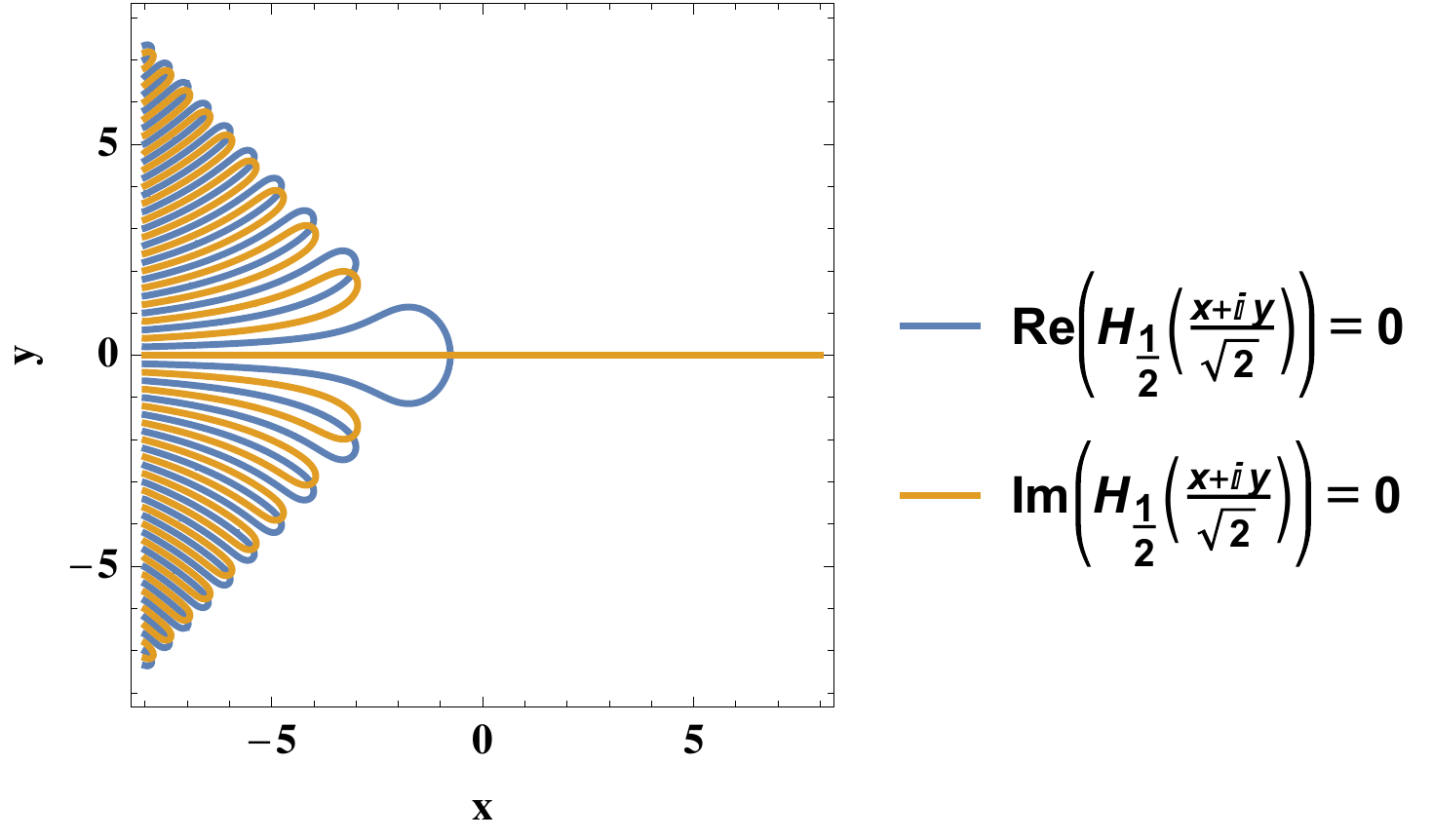}
\end{center}
\vspace{-.5cm}
\caption{Lines on the complex plane $z=x+iy$ where the real and imaginary parts of $H_{\frac{1}{2}}\left(\frac{z}{\sqrt{2}}\right)$  are zero.}
\label{PolesReImZeroLinesN5}
\end{figure}

\begin{figure}[h]
\begin{center}
\includegraphics[width=5.3cm]{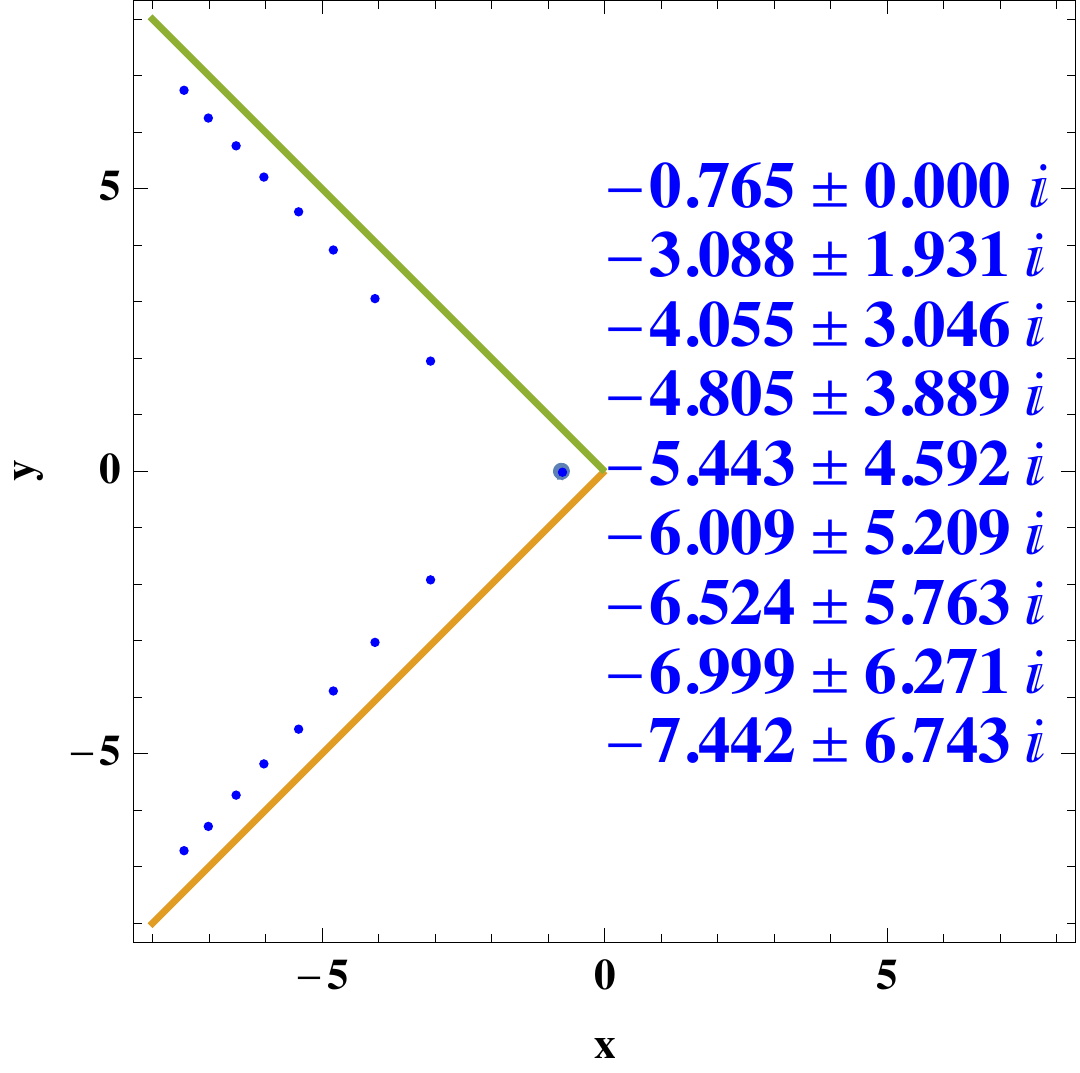}
\end{center}
\vspace{-.5cm}
\caption{The 17 zeroes on the complex plane $z=x+iy$ for $H_{\frac{1}{2}}\left(\frac{z}{\sqrt{2}}\right)$ with the largest real part.}
\label{PolesZeroesN5}
\end{figure}

There is no need to discuss the  singularity structure of the $N=4$  amplitude. In the $N=5$ case the partial wave has  a set of simple poles situated at the zeroes of the 
function $H_{1 \over 2} \left({z \over \sqrt{2}}\right)$. The lines where the real and imaginary parts of this function vanish are shown in Fig.~\ref{PolesReImZeroLinesN5}. 
The poles are placed at the points where these lines intersect and are plotted, together with their numerical values, in Fig.~\ref{PolesZeroesN5}.
The two arrays of poles are situated just below the line with argument equal to $ \frac{3\pi}{4}$ and above the line with argument equal to $\frac{5 \pi}{4}$.  

The simple structure of the $N=6$ case stems from being the solution to
\begin{eqnarray}
\ f_\omega^{(6)}  ~= 1- \alpha\,  t \frac{d}{d\,\omega}\,\frac{f_\omega^{(6)} }{\omega} \, ,
\end{eqnarray}
since $\eta_6 =0$.  The partial wave can be written as
\begin{eqnarray}
f_\omega^{(6)}  &=& \frac{\sqrt{2} w }{\sqrt{- \alpha t}}
H_{-1}\left(\frac{w}{\sqrt{-2 \alpha t}}\right)
~ \sim ~ \sum_{n=0}^\infty \frac{(2n)!}{n!} \left(\frac{\alpha t}{2 \omega^2}\right)^n
\end{eqnarray}
where ${\cal C}_n^{(6)} =   \frac{(2n)!}{2^n n!}$. The analytic structure in the complex $\omega$ plane is therefore trivial. 

\begin{figure}[h]
\begin{center}
\includegraphics[width=9cm]{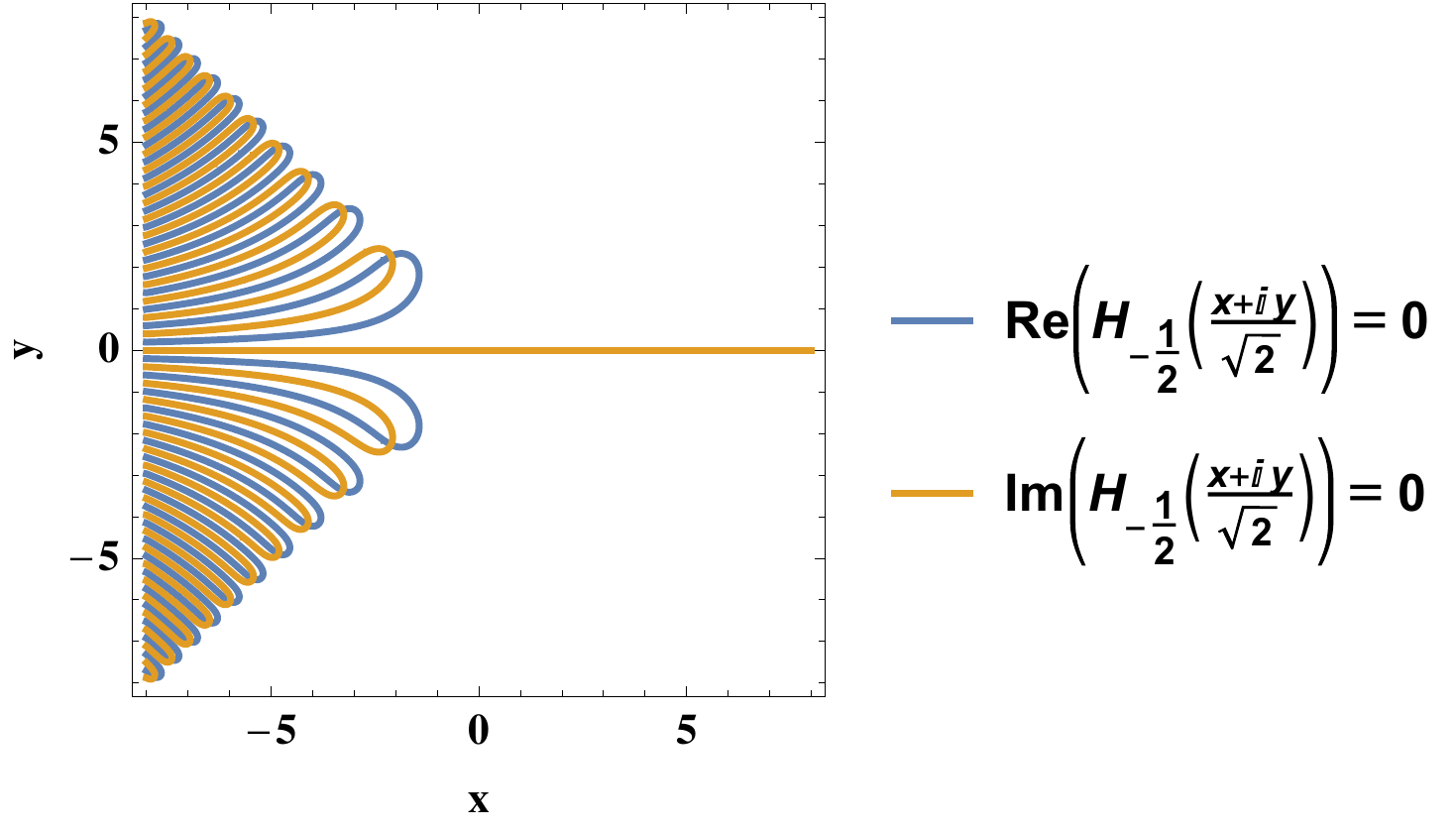}
\end{center}
\vspace{-.5cm}
\caption{Lines on the complex plane $z=x+iy$ where the real and imaginary parts of $H_{-\frac{1}{2}}\left(\frac{z}{\sqrt{2}}\right)$  are zero.}
\label{PolesZeroLinesN7}
\end{figure}

\begin{figure}[h]
\begin{center}
\includegraphics[width=9cm]{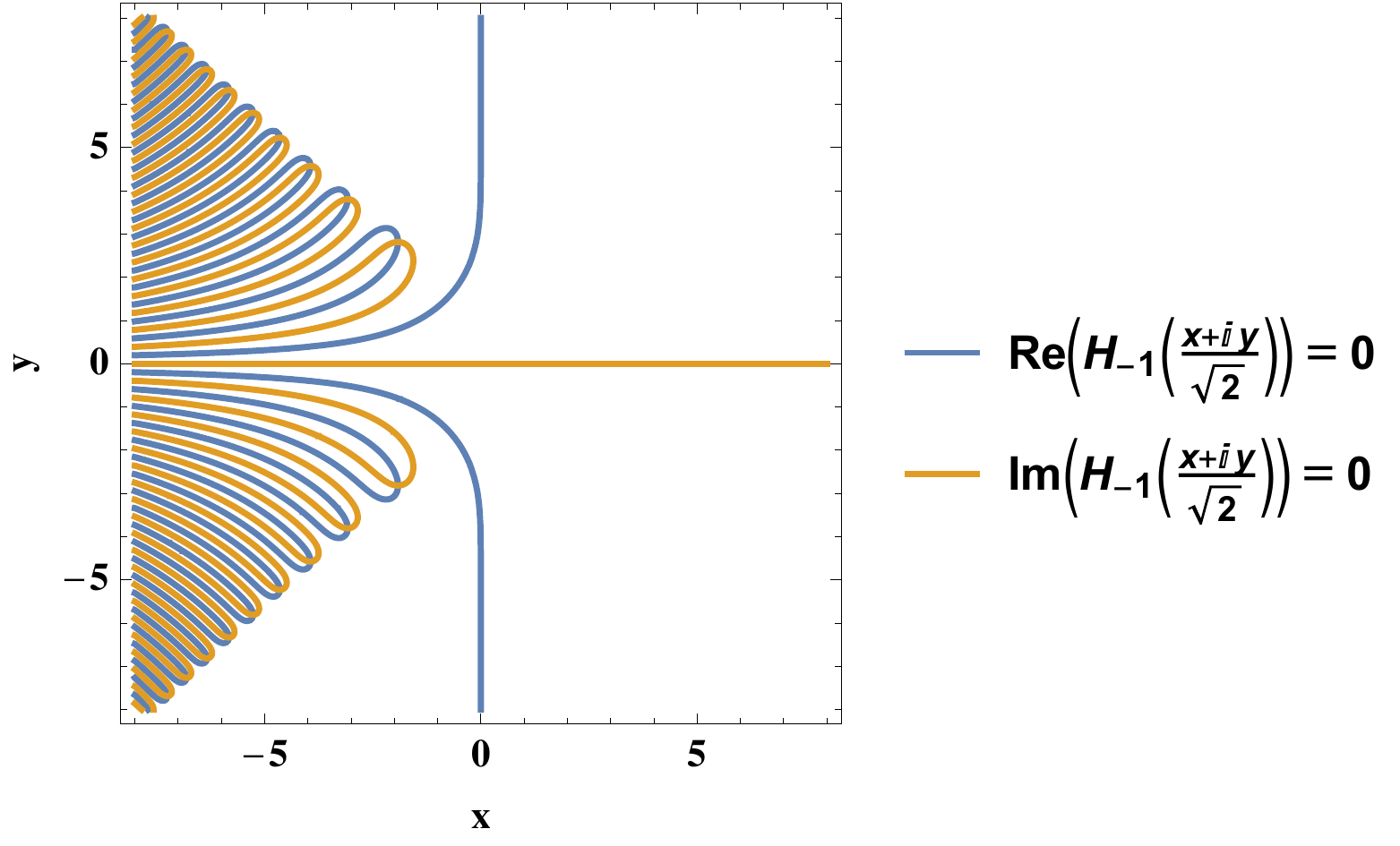}
\end{center}
\vspace{-.5cm}
\caption{Lines on the complex plane $z=x+iy$ where the real and imaginary parts of $H_{-1}\left(\frac{z}{\sqrt{2}}\right)$  are zero.}
\label{PolesZeroLinesN8}
\end{figure}

\begin{figure}[h]
\begin{center}
\includegraphics[width=5.3cm]{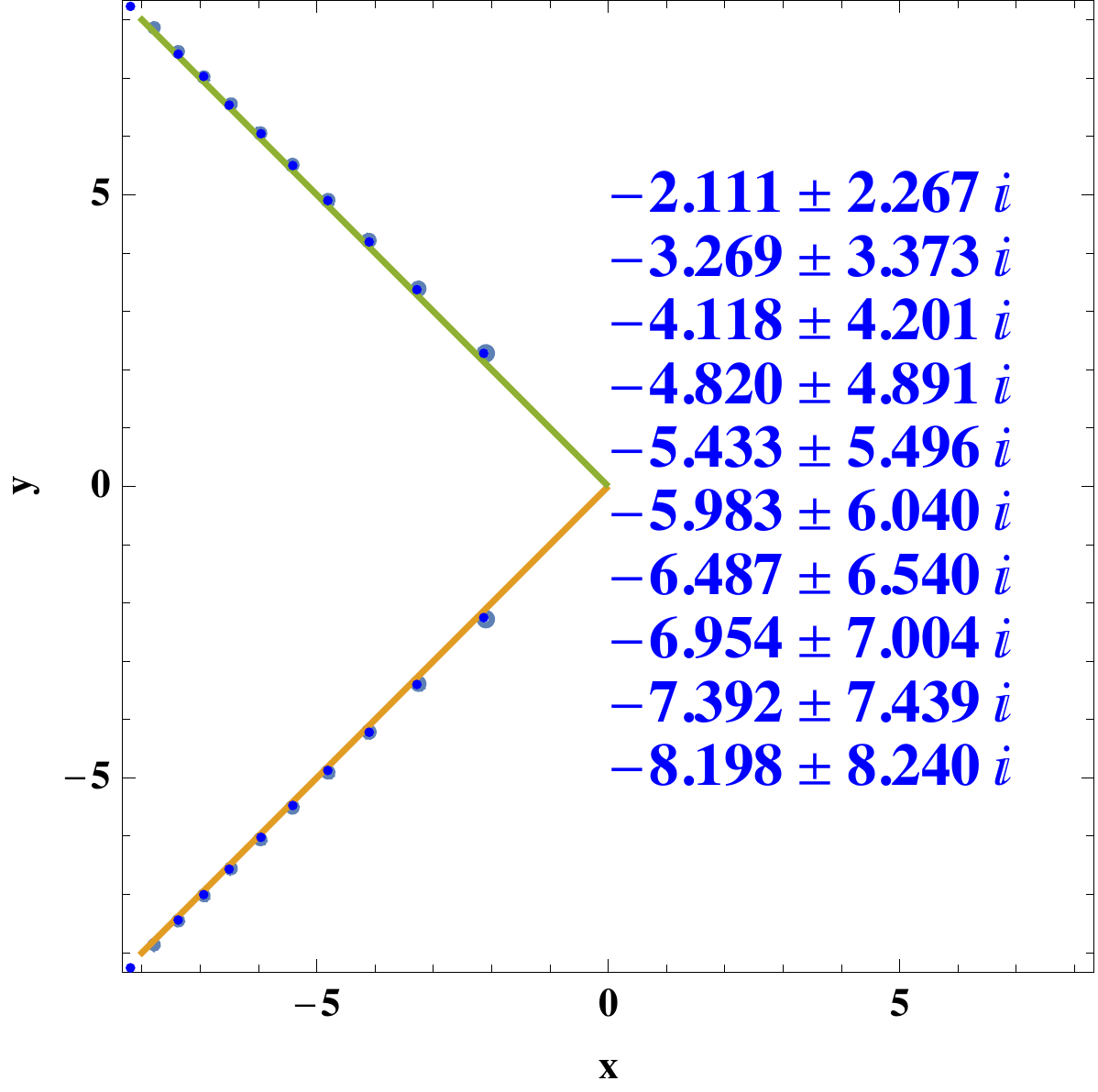}
\end{center}
\vspace{-.5cm}
\caption{The 20 zeroes on the complex plane $z=x+iy$ for $H_{-\frac{1}{2}}\left(\frac{z}{\sqrt{2}}\right)$ with the largest real part.}
\label{PolesZeroesN7}
\end{figure}

\begin{figure}[h]
\begin{center}
\includegraphics[width=5.3cm]{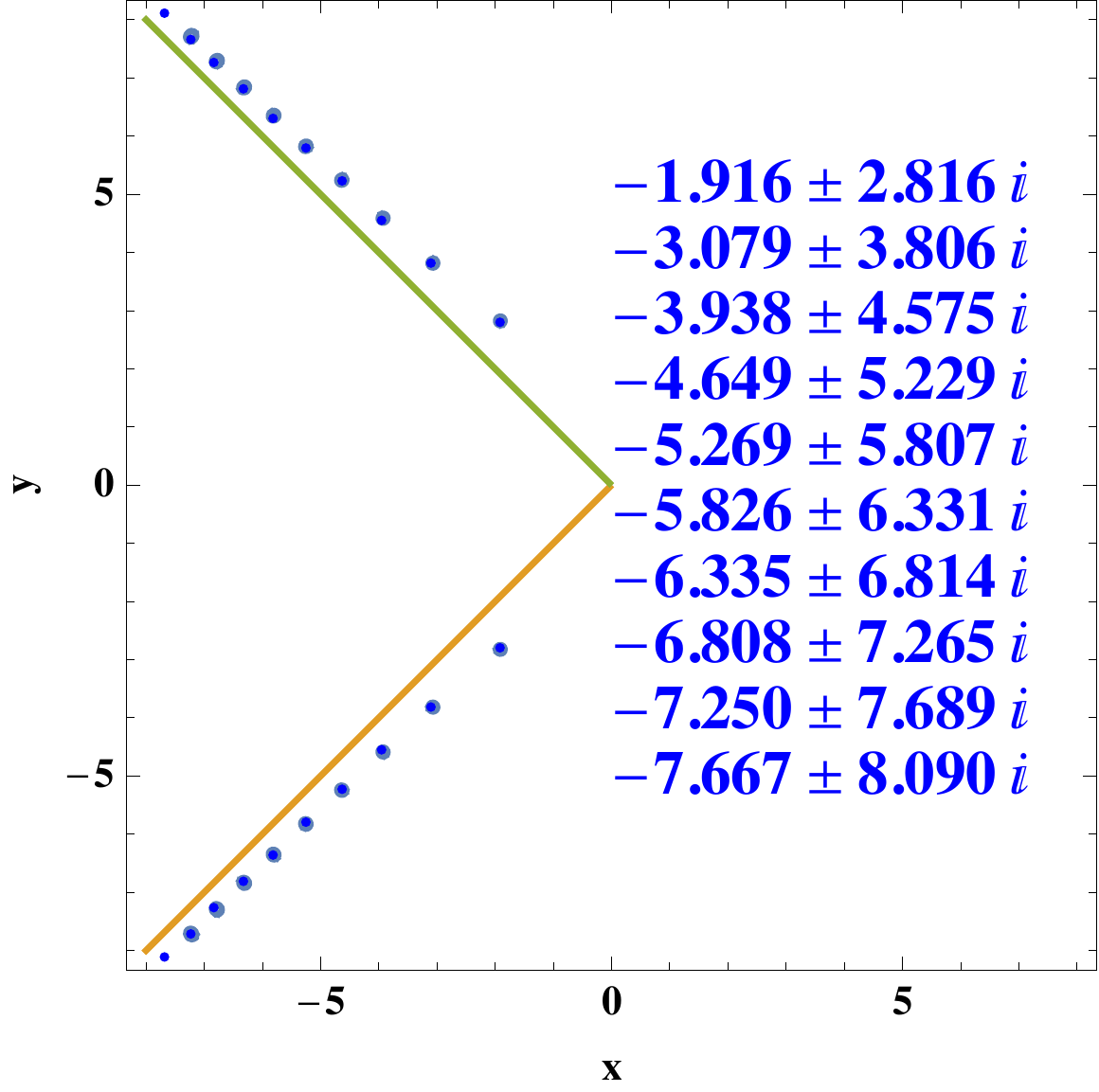}
\end{center}
\vspace{-.5cm}
\caption{The 20 zeroes on the complex plane $z=x+iy$ for $H_{-1}\left(\frac{z}{\sqrt{2}}\right)$ with the largest real part.}
\label{PolesZeroesN8}
\end{figure}

The location of the partial wave singularities is  at the zeroes of 
$H_{- 1 \over 2} \left({z \over \sqrt{2}}\right)$ for $N=7$ and $H_{-1} \left({z \over \sqrt{2}}\right)$ for $N=8$. These functions have their lines with zero real and imaginary parts shown in Fig.~\ref{PolesZeroLinesN7} and Fig.~\ref{PolesZeroLinesN8}. Their zeroes with the largest real parts, and correspondingly, the singular points for the partial waves, together with their numerical values are given in Fig.~\ref{PolesZeroesN7} and Fig.~\ref{PolesZeroesN8}.
These simple poles are now situated close and above the line with argument equal to $ \frac{3\pi}{4}$ and below the line with argument equal to $\frac{5 \pi}{4}$ as shown in Fig.~\ref{PolesZeroesN7} and Fig.~\ref{PolesZeroesN8}.

\begin{figure}
\begin{center}
\includegraphics[width=8cm]{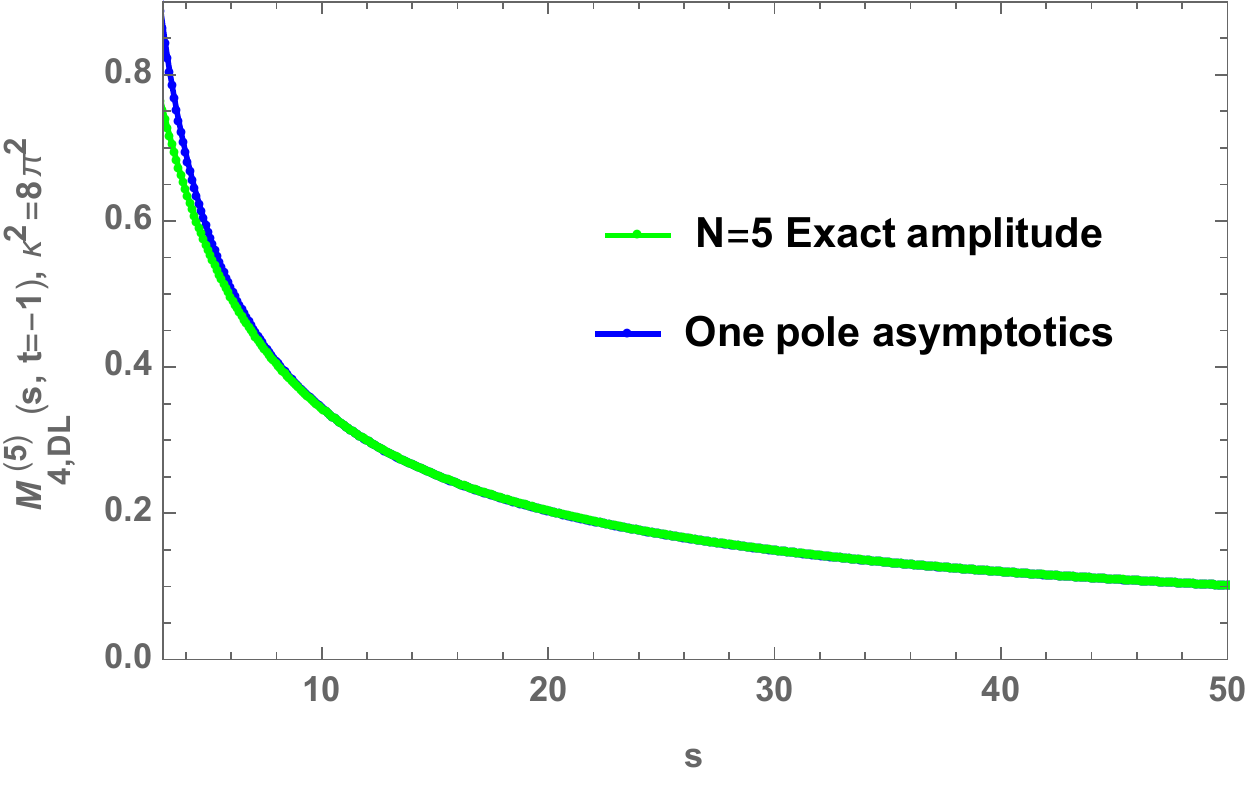}
\includegraphics[width=8cm]{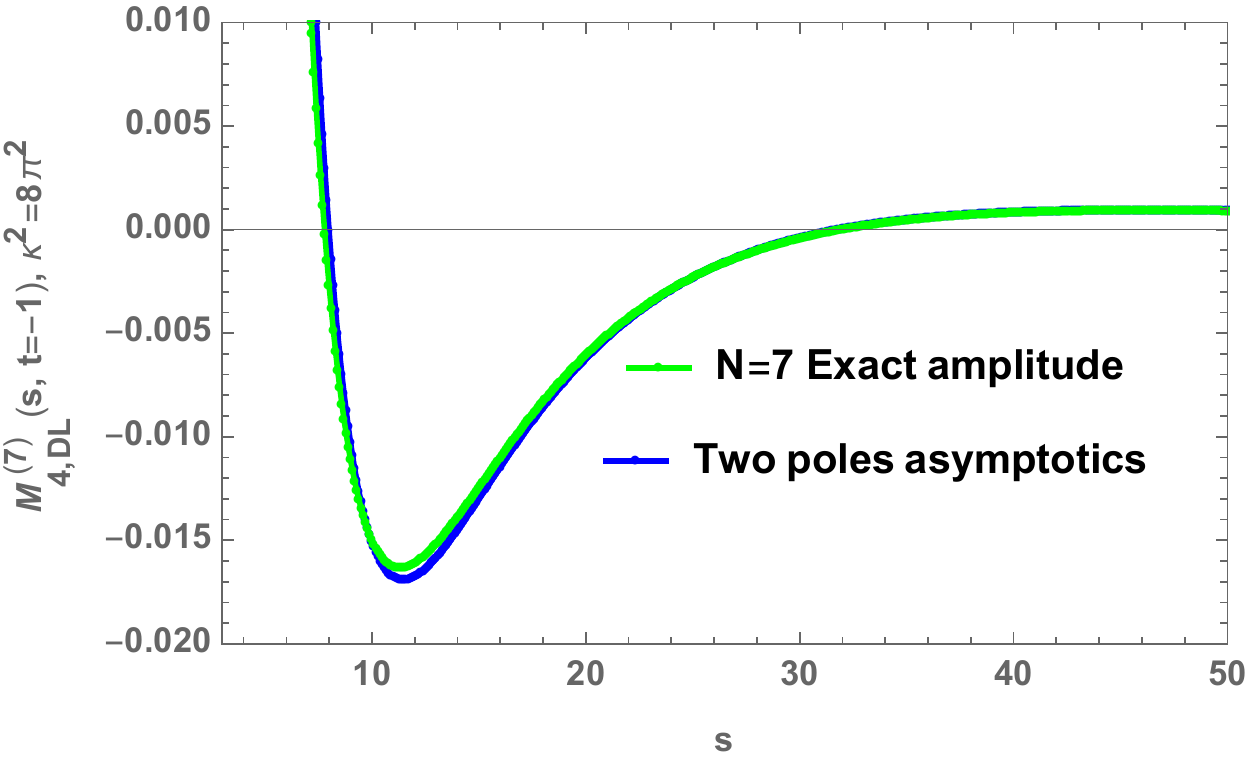}
\includegraphics[width=8cm]{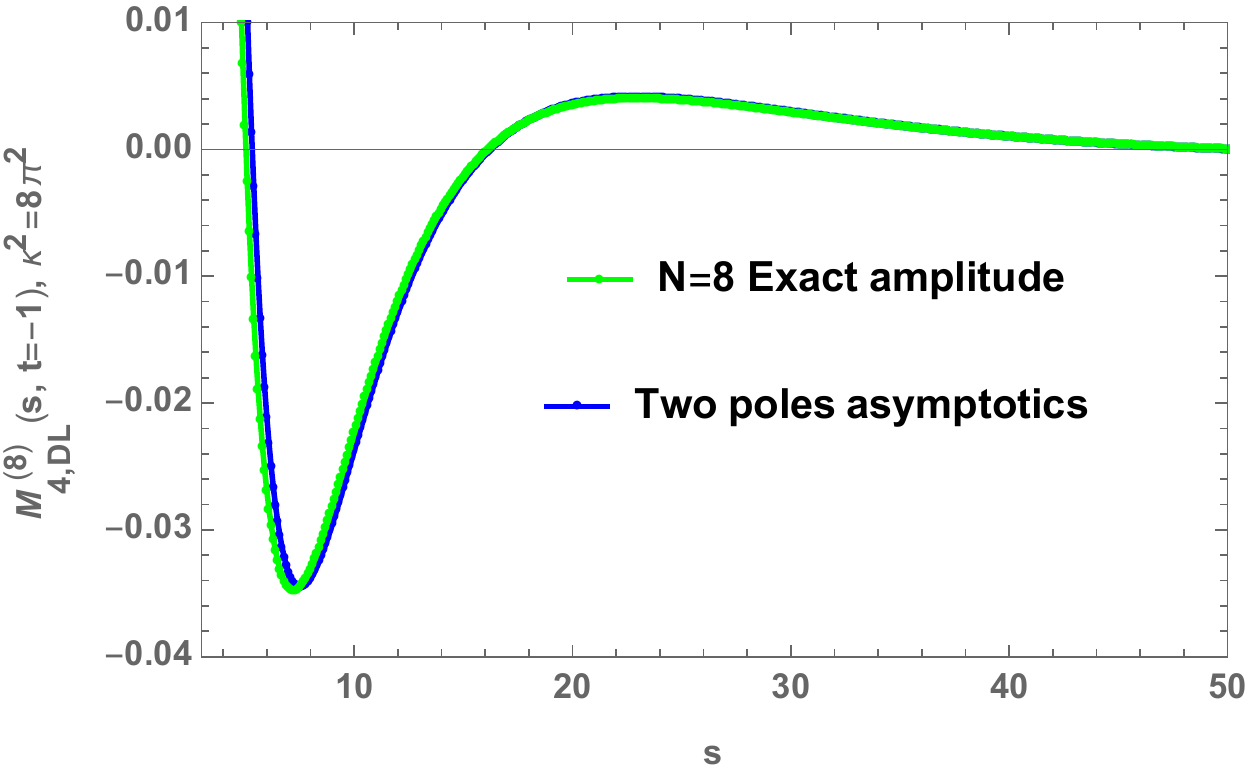}
\vspace{-.5cm}
\end{center}
\caption{A comparison of the DL exact amplitudes with their asymptotic estimates for different supergravities with $N=5,7,8$ as a function of $s$ with $\kappa^2 = 8 \pi^2$ and $t = -1$. }
\label{ExactAsympNAmp}
\end{figure}

The $\frac{s}{-t} \to \infty$ asymptotic expressions for the different DL amplitudes considering only the rightmost poles 
in each case can be written in the form (the exact expression is given for the  $N=6$ amplitude)
 \begin{eqnarray}
 {\cal M}_{4,{\rm DL}}^{(5)}  (s,t)
&\sim& 2 \left(\frac{s}{-t}\right)^{-0.765 \sqrt{- \alpha t}} \, , 
\label{M4N5}\\
{\cal M}_{4,{\rm DL}}^{(6)}  (s,t)
&=& e^{\frac{\alpha}{2} t \ln^{2}{\left(\frac{s}{-t}\right)}} \, , 
\label{M4N6}\\
{\cal M}_{4,{\rm DL}}^{(7)}  (s,t)
&\sim& -4 \left(\frac{s}{-t}\right)^{-2.111 \sqrt{- \alpha t}} \cos{\left(2.267 \sqrt{- \alpha\, t} 
\ln{\left(\frac{s}{-t}\right)}\right)} \,  , 
\label{M4N7}\\
{\cal M}_{4,{\rm DL}}^{(8)}  (s,t)
&\sim& - 2 \left(\frac{s}{-t}\right)^{-1.916 \sqrt{- \alpha t}} \cos{\left(2.816 \sqrt{- \alpha\, t} 
\ln{\left(\frac{s}{-t}\right)}\right)} \,  .
\label{M4N8}
\end{eqnarray}
The first factor in these expressions stems from the evaluation of the residues which generates a term $2/(6-N)$. The comparison of these asymptotic results, for $\kappa^2 = 8 \pi^2$ and $t = -1$, with their exact counterparts is shown in Fig.~\ref{ExactAsympNAmp}. 

\section{Impact parameter representation}

To better understand the physical content of the DL resummation it is useful 
to work in impact parameter space, $\rho$, for the graviton--graviton interaction. 
The required Fourier transformation of the amplitude reads
\begin{eqnarray}
\chi^{(N)} \left(\rho,s\right) &=&  \alpha s 
\int \frac{d^2 \vec{q}}{q^2} e^{- i \vec{q} \cdot \vec{\rho}} 
{\cal M}_{4,{\rm DL}}^{(N)}  (s,t) \, ,
\label{phaseikonal} 
\end{eqnarray}
where $q=\sqrt{-t}$. In a high energy and fixed impact parameter eikonal approach it is now possible to exponentiate this phase, 
\begin{eqnarray}
\frac{\alpha s^{2}}{q^{2}} {\cal M}_{4,{\rm DL}}^{(N)}  (s,t) 
~=~ \frac{s}{4 \pi^{2}} \int d^{2} \vec{\rho} \, e^{i \vec{q} \cdot \vec{\rho}} \chi^{(N)} (\rho, s) 
~\simeq~ 
-\frac{i s}{4 \pi^{2}} \int d^{2} \vec{\rho} \, e^{i \vec{q} \cdot \vec{\rho}}\left(e^{i \chi^{(N)} (\rho, s)}-1\right). 
\end{eqnarray}
Since $s>0$ and $t=-\frac{s}{2}(1-\cos \theta) \leq 0$, the $s \gg -t$ limit implies $t=-q^{2} \simeq -s \frac{\theta^{2}}{4}$ and, hence, $\theta \ll 1$. The integration above is dominated by the stationary phase region given by $ \frac{\partial}{\partial \rho}\left(q \rho+\chi^{(N)}\right)=0$. It is then possible to evaluate the associated graviton--graviton scattering angle in the DL eikonal 
and forward limit taking the derivative with respect to $\rho$,
\begin{eqnarray}
\theta^{(N)} (\rho,s) &=& -\frac{2}{\sqrt{s}}\frac{\partial 
\chi^{(N)} \left(\rho,s\right)}{\partial \rho}   \, .
\end{eqnarray}

The simplest case for study 
is that of $N=4$ since ${\cal M}_{4,{\rm DL}}^{(4)} =1$ and  its eikonal phase corresponds to the impact parameter representation of the Born term in 
Eq.~(\ref{eqnnotation}) evaluated at high energies, where $u\simeq -s$, and fixed $t=-q^2$. It corresponds to a logarithm containing the infrared scale $\lambda$,
\begin{eqnarray}
\chi^{(4)} \left(\rho,s\right) &=& 
 \frac{1 }{2 s}
\int \frac{d^2 \vec{q}}{(2 \pi)^2} e^{-i \vec{q} \cdot \vec{\rho}} 
{\cal A}_4^{\rm Born} (s,t) ~\simeq~
 \frac{\kappa^2 s }{8 \pi^2}
\int \frac{d^2 \vec{q}}{q^2} e^{-i \vec{q} \cdot \vec{\rho}} \, \theta\left(q - \lambda \right)
\nonumber\\
&\Rightarrow& - \frac{\kappa^2 s}{4 \pi} \ln{\left(\rho \, \lambda\right)} 
~=~ \chi_{\rm Born} (\rho,s) \, .
\end{eqnarray}
The symbol ``$\Rightarrow$" has been introduced to indicate the limit when the cut--off $\lambda$ goes to zero.  The associated graviton's deflection angle in the forward limit, which is $\lambda$ independent, then reads
\begin{eqnarray}
\theta^{(4)}(\rho,s)   ~ = ~    \frac{\kappa^2 \sqrt{s}}{2 \pi  \rho}  
~=~ \theta_{\rm Born} (\rho,s) \, .
\end{eqnarray}
For a given $s$, the growth of $\theta_{\rm Born}$ in the limit $\rho \to 0$ is characteristic of an attractive interaction. 

The Born term can be subtracted in order to isolate the DL effects in those theories with $N>4$. The DLs are subject to the constraint $s > -t$~\footnote{This is a different approach to the asymptotic, and less accurate, one presented in~\cite{SabioVera:2019edr} for the $N=8$ case. Since the DL resummation is valid in the $s \gg -t = q^2$ limit it is natural to constrain the integration in the eikonal phase to the region where $q < \sqrt{s}$. A second improvement is to consider the $t$ dependence in the argument of the logarithms, which was treated as a subleading effect in~\cite{SabioVera:2019edr}. This allows for a better description of the interplay between impact parameter and energy in the quantities here investigated. }, {\it i.e.}
 \begin{eqnarray}
\chi^{(N)} \left(\rho,s\right) &=&   \chi_{\rm Born} (\rho,s)  
+ \chi_{\rm DL}^{(N)} \left(\rho,s\right)  \, ,
\end{eqnarray} 
where (see Eq.~(\ref{PertAmplDL4}))
 \begin{eqnarray}
\chi_{\rm DL}^{(N)} \left(\rho,s\right) &=&   \alpha s 
\int \frac{d^2 \vec{q}}{q^2} e^{- i \vec{q} \cdot \vec{\rho}} 
 \sum_{n=1}^\infty {\cal C}_n^{(N)} \, 
\frac{(\alpha \, t)^n}{(2n)!} \ln^{2n}{\left(\frac{s}{-t}\right)} \,  \theta \left(\sqrt{s}-q\right)  
\nonumber\\
 &=&      \pi
\sum_{m=1}^\infty    \sum_{n=0}^{m-1} 
\frac{ (-1)^{m} (\alpha s)^{m+1} {\cal C}_{m-n}^{(N)}}{(n!)^2 m^{2 (m-n) +1}}   
\left({\rho^2 \over 4 \alpha}\right)^{n}  \, .
\label{chidelcoeffs}
\end{eqnarray} 

In this DL sector of the eikonal phase the regulator $\lambda$ can be dropped out since the integration is free from infrared singularities. When expanded in powers of the coupling, this representation of the eikonal phase correctly reproduces the eikonal--DL crossed terms of the form
\begin{eqnarray}
\sim  i \pi s^3  \frac{(N-4)}{12} 
\ln^3{\left(\frac{s}{-t}\right)}
\end{eqnarray}
present in the two--loop exact four--graviton scattering amplitudes for different supergravities~\cite{BoucherVeronneau:2011qv}. 

\begin{figure}[h]
\begin{center}
\includegraphics[width=9cm]{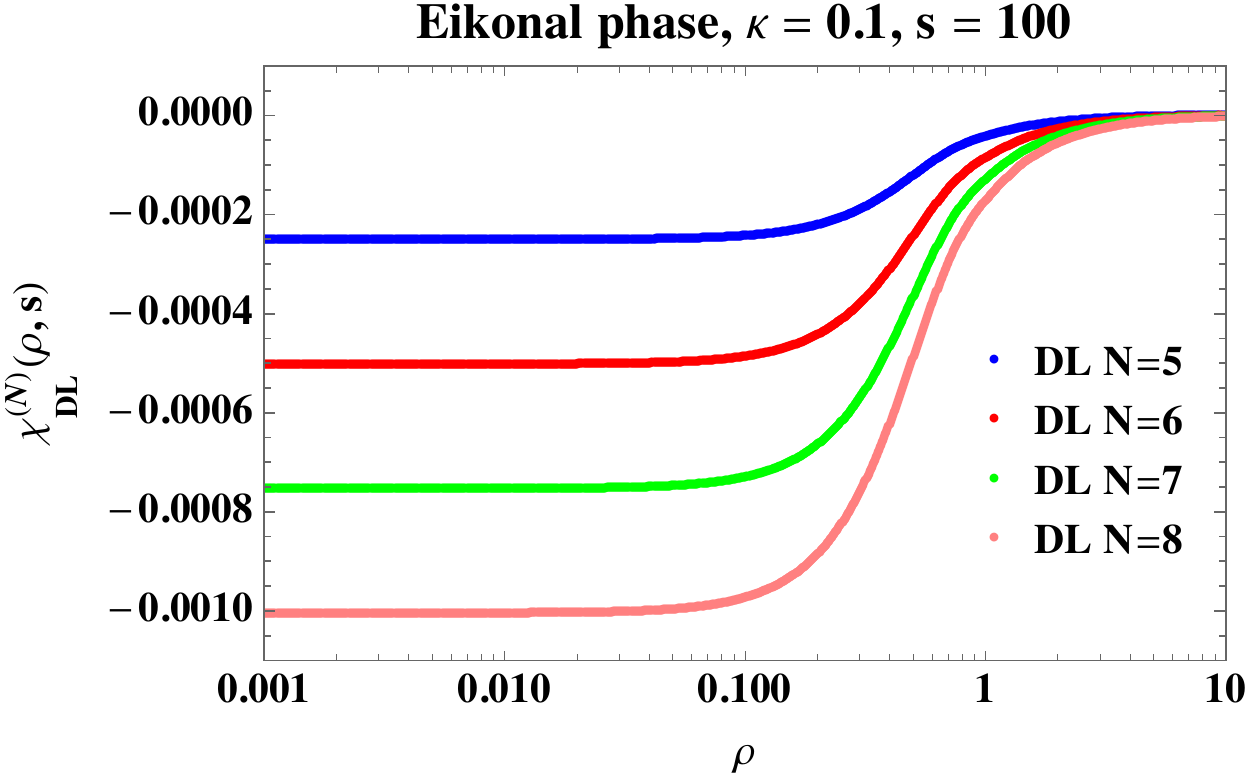}
\end{center}
\vspace{-.5cm}
\caption{Contribution to the eikonal phase from DLs in 
different supergravities as a function of 
the impact parameter $\rho$ with $\kappa = 0.1$ and 
$s = 100$.}
\label{ChiSeveralN}
\end{figure}

\begin{figure}[h]
\begin{center}
\includegraphics[width=9cm]{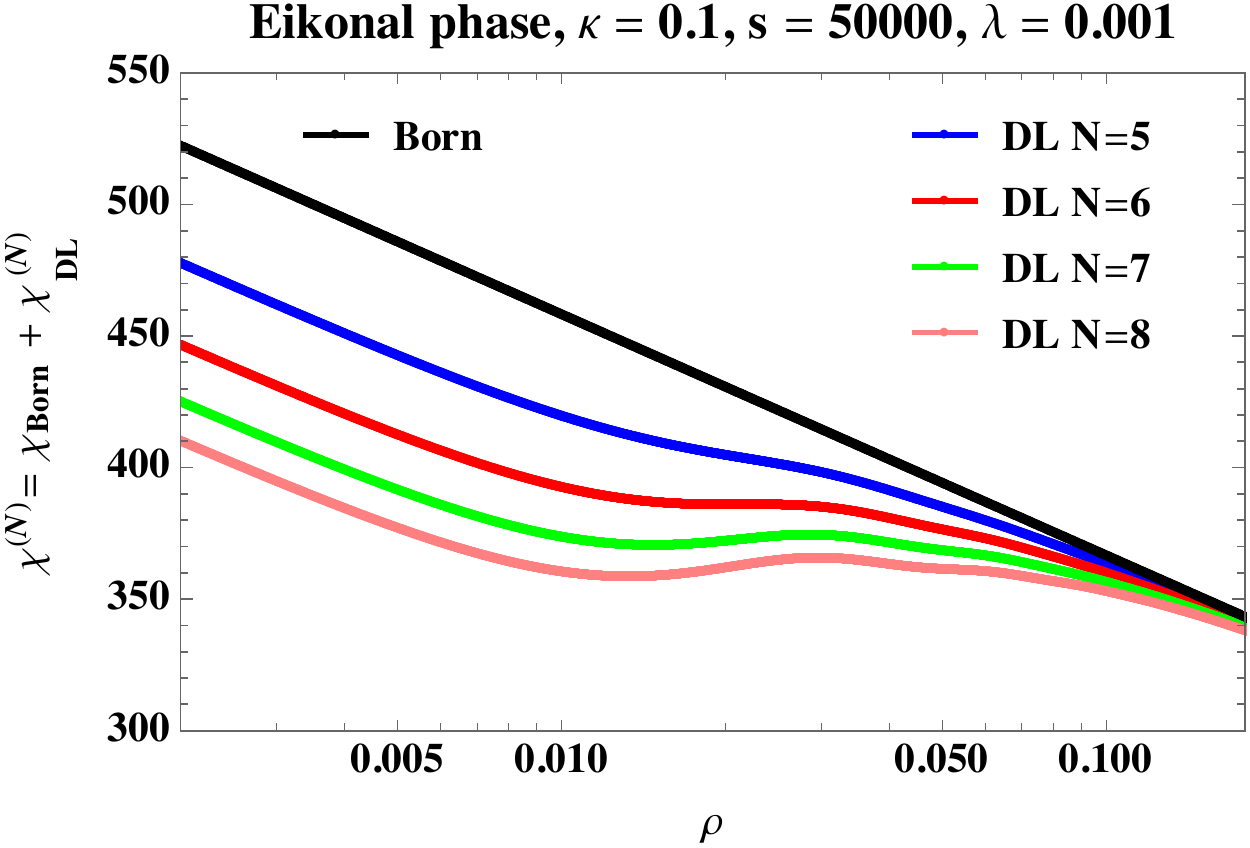}
\end{center}
\vspace{-.5cm}
\caption{The Born eikonal phase and its value plus the  DL corrections for a fixed energy $s=50000$ and different supergravity theories. }
\label{EikonalPhaseSeveralN}
\end{figure}

A numerical study of the contribution to the eikonal phase from DLs  as a function of the impact parameter $\rho$ with $\kappa = 0.1$ and 
$s = 100$ is given in Fig.~\ref{ChiSeveralN}. For all ${\cal N}>4$ supergravities this is a negative function which goes rapidly to zero as the impact parameter grows. Therefore at large $\rho$ the Born  phase prevails and the DLs are only relevant at 
small distances. The modification of the Born phase is more evident as the number of gravitinos in the theory increases. The equidistance among different lines arises because 
in the $\rho \to 0$ limit with not very large energies the role of the DLs is to generate a constant shift in the eikonal phase of the form
 \begin{eqnarray}
\chi^{(N)} \left(\rho,s\right) &\overset{ \rho \ll 1}{\simeq}&   \chi_{\rm Born} (\rho,s)  
+ \left(4-N\right) \frac{\kappa^4 s^2}{128 \pi^3}  \, ,
\label{constantphaseshift}
\end{eqnarray} 
which stems from the coefficient ${\cal C}_{1}^{(N)}$. For any value of $s$ the role of the higher order DLs is to suppress this shift in the phase at larger values of the impact parameter and generate a consistent Born behaviour at long distances. Hence the  phase shift in this region  is governed by the one--loop DL corrections, it is proportional to $\kappa^4 s^2$ and with opposite sign to the Born eikonal phase which goes as $\kappa^2 s$. This 
implies that the relative size of the DLs to the Born phase rapidly grows with $s$.  To further understand this point, the full eikonal phase, with both the Born and DL terms included, is shown in Fig.~\ref{EikonalPhaseSeveralN} for a large center of mass energy $s=50000$. It can again be seen how the DL resummation generates a constant phase shift which is more pronounced as $N$ grows. In the region with very high energies Eq.~(\ref{constantphaseshift}) is no longer an accurate approximation and a large number of coefficients are needed to find the eikonal phase shift for $\rho \ll 1$.  

The deflection angle with DLs for each supergravity  can be written in the form
\begin{eqnarray}
\theta^{(N)} \left(\rho,s\right)  &=& \theta_{\rm Born} \left(\rho,s\right) +
\theta_{\rm DL}^{(N)} \left(\rho,s\right) 
\end{eqnarray}
where
\begin{eqnarray}
\theta_{\rm DL}^{(N)} \left(\rho,s\right) 
=   - \theta_{\rm Born} \left(\rho,s\right)  
\sum_{m=2}^\infty    \sum_{n=1}^{m-1} 
\frac{ n  (- \alpha s)^{m} {\cal C}_{m-n}^{(N)}}{(n!)^2 m^{2 (m-n) +1}}    
 \left({\rho^2 \over 4 \alpha}\right)^{n} 
 \, .
 \label{ThetaDLN}
\end{eqnarray} 
\begin{figure}[h]
\begin{center}
\includegraphics[width=9cm]{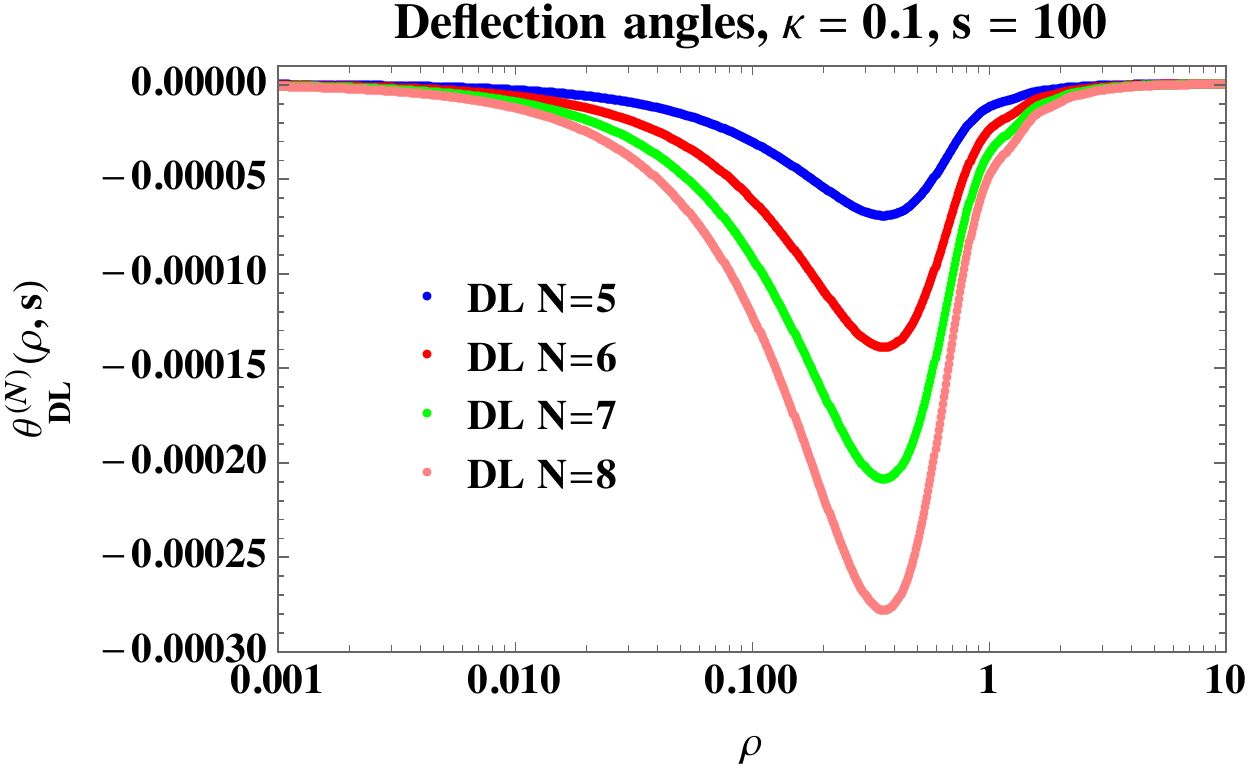}
\end{center}
\vspace{-.5cm}
\caption{Contribution to the deflection angle stemming from DLs in different supergravities as a function of 
the impact parameter $\rho$ with $\kappa = 0.1$ and 
$s = 100$.}
\label{ThetaSeveralN}
\end{figure}

\begin{figure}[h]
\begin{center}
\includegraphics[width=9cm]{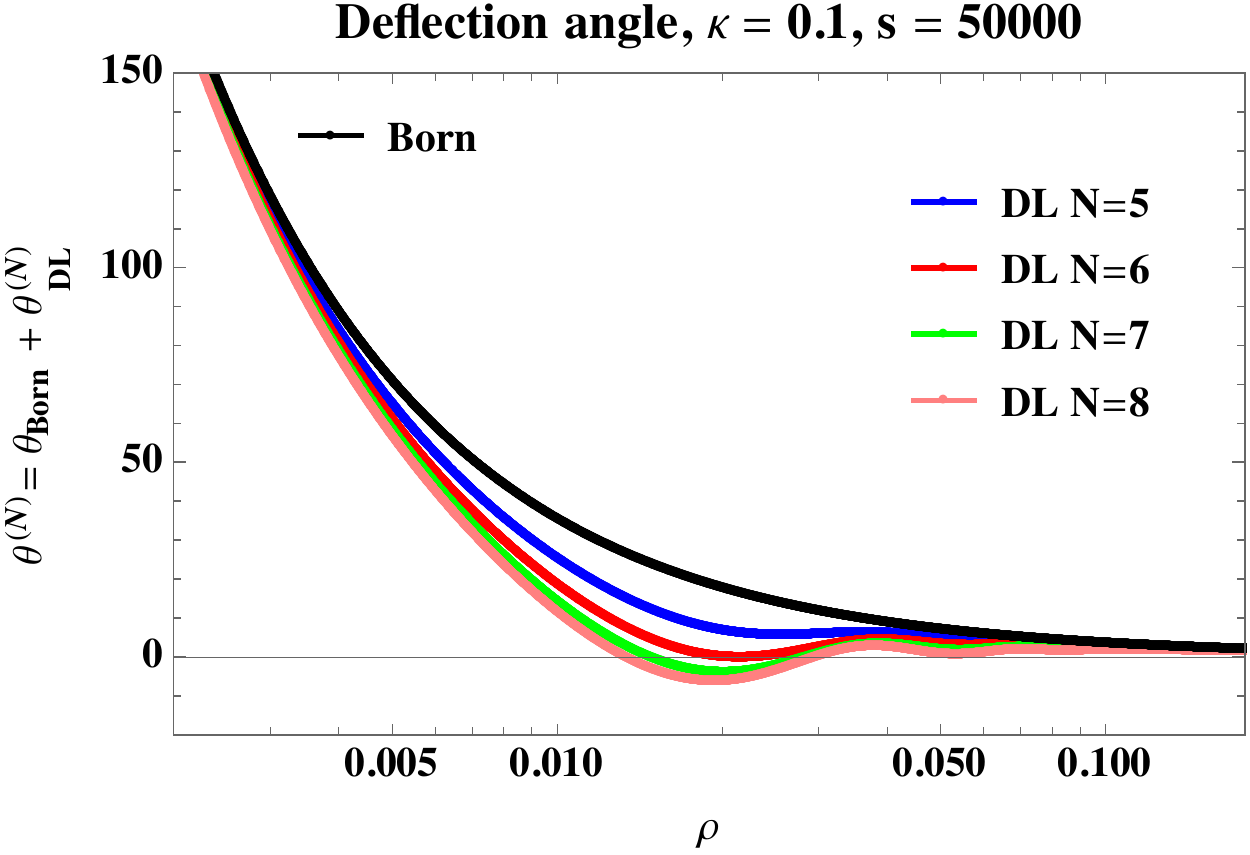}
\end{center}
\vspace{-.5cm}
\caption{Graviton's deflection angle for a large value of the center of mass energy as a function of the impact parameter space with and without DL corrections for $N=5,6,7,8$ supergravity theories.}
\label{DeflectionAngleSeveralN}
\end{figure}

A plot of the DLs to this angle, $\theta_{\rm DL}^{(N)}$, is shown for different $N$ and a moderate value of $s$ in Fig.~\ref{ThetaSeveralN}. Since $\theta_{\rm DL}^{(N)}$ is always negative, the DL resummation has the effect of reducing the Born graviton--graviton deflection angle. This means that a screening of the gravitational interaction is generated by the DL quantum corrections to the scattering amplitude which is larger as the number of gravitinos in the theory increases. The same effect was already found in an asymptotic analysis for ${\cal N}=8$ supergravity in~\cite{SabioVera:2019edr}. The screening of gravity is stronger as $s$ grows, up to the point that it can generate a negative graviton's  deflection angle for high enough center of mass energies. To illustrate this 
point Fig.~\ref{DeflectionAngleSeveralN} is given for a very large $s$ where the angle is evaluated at Born level and including the DLs for $N=5,6,7,8$. It can be seen that $\theta^{(N)}<0$ for a large region in impact parameter space. 
The higher the number of gravitinos in the theory, the more important this weakening of gravity effect becomes. 

The DLs increase their relative size with respect to the Born reflection angle and their range of action in  impact parameter space  as the center--of--mass energy grows.  The modification of the Born behaviour at small distances for $N=8$ is the largest of all the supergravities here discussed. For each theory there is an impact parameter  $\rho_{\rm max}^{(N)}$ where 
the corrections to the deflection angle peak. In order to get its exact value all the DL terms should be taken into account and it is needed to solve the equation
\begin{eqnarray} 
\sum_{m=1}^\infty    \sum_{n=0}^{m-1} 
\frac{ (- \alpha s)^{m} {\cal C}_{m-n}^{(N)}}{(n!)^2 (m+1)^{2 (m-n) +1}}    
 \left({ {\rho_{\rm max}^{(N)}}^2 
 \over 4 \alpha}\right)^{n} &=& 0 \, .
\end{eqnarray} 

\begin{figure}[h]
\begin{center}
\includegraphics[width=9cm]{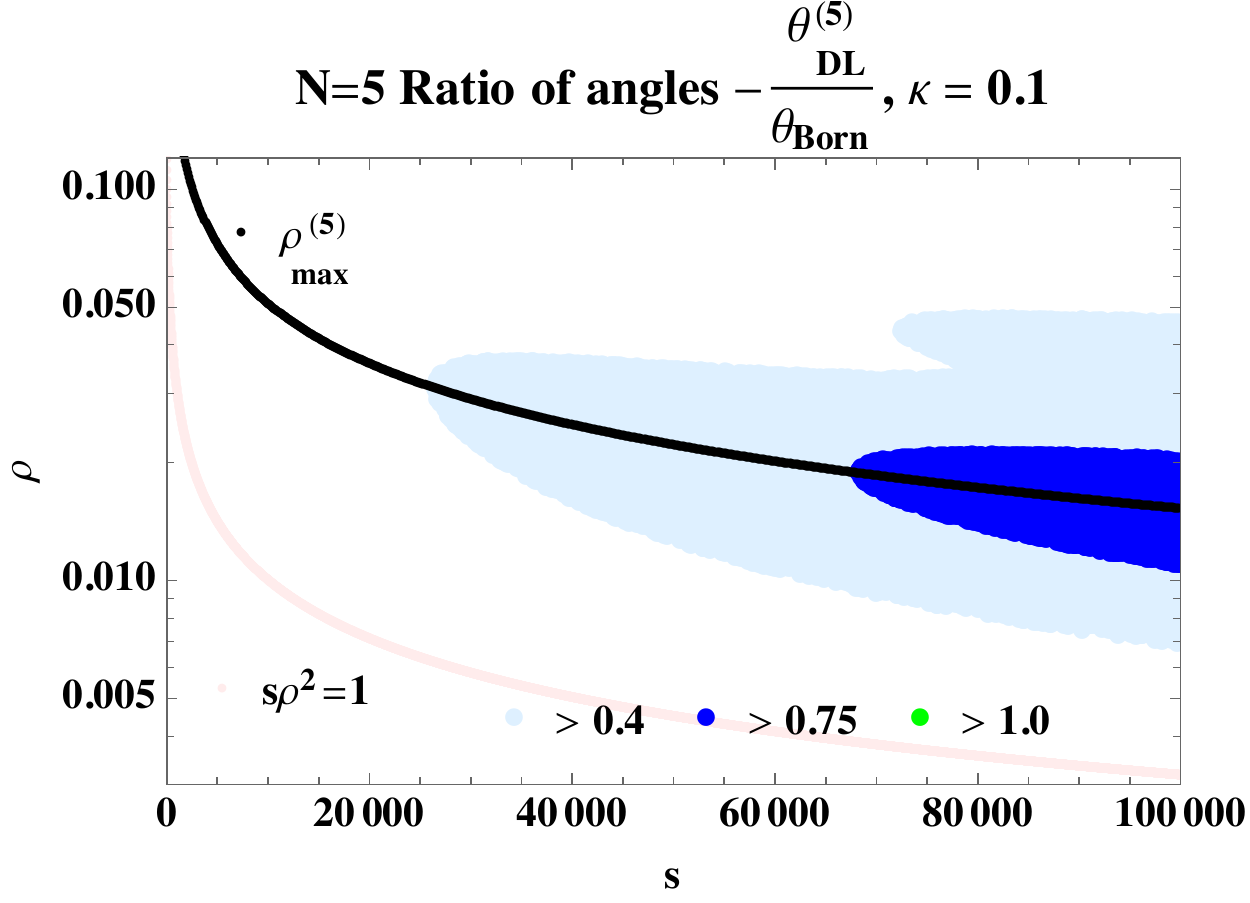}
\includegraphics[width=9cm]{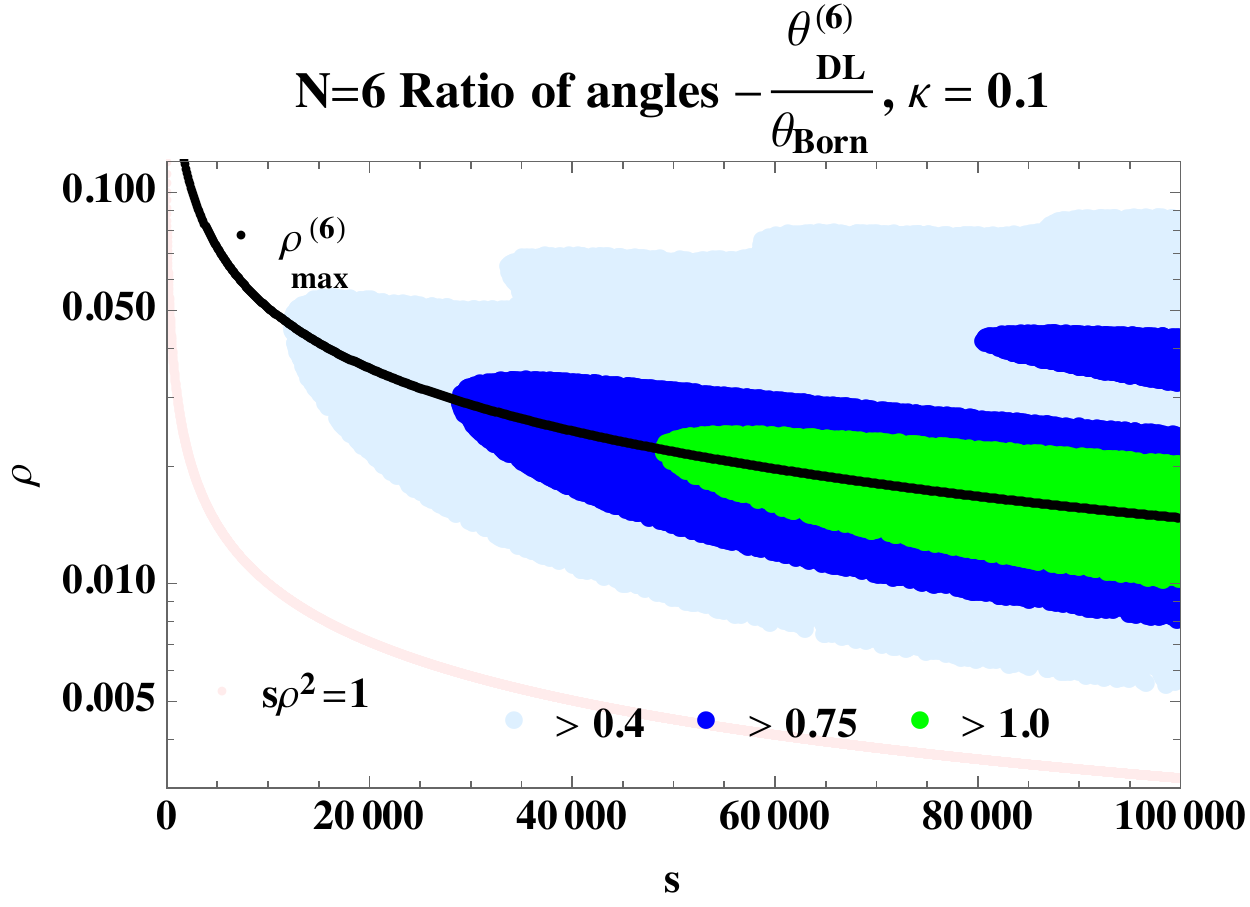}
\end{center}
\vspace{-.5cm}
\caption{Regions in impact parameter and energy space where the DL resummation effects are large ($N$=5 up, $N$=6 bottom). When the ratio $-\frac{\theta^{(N)}_{\rm DL}}{\theta_{\rm Born}}>1$ the deflection angle is negative. When it is bigger than 0.4 and 0.75, it corresponds to a large correction but still with a positive angle. }
\label{N56RatioAngles}
\end{figure}

\begin{figure}[h]
\begin{center}
\includegraphics[width=9cm]{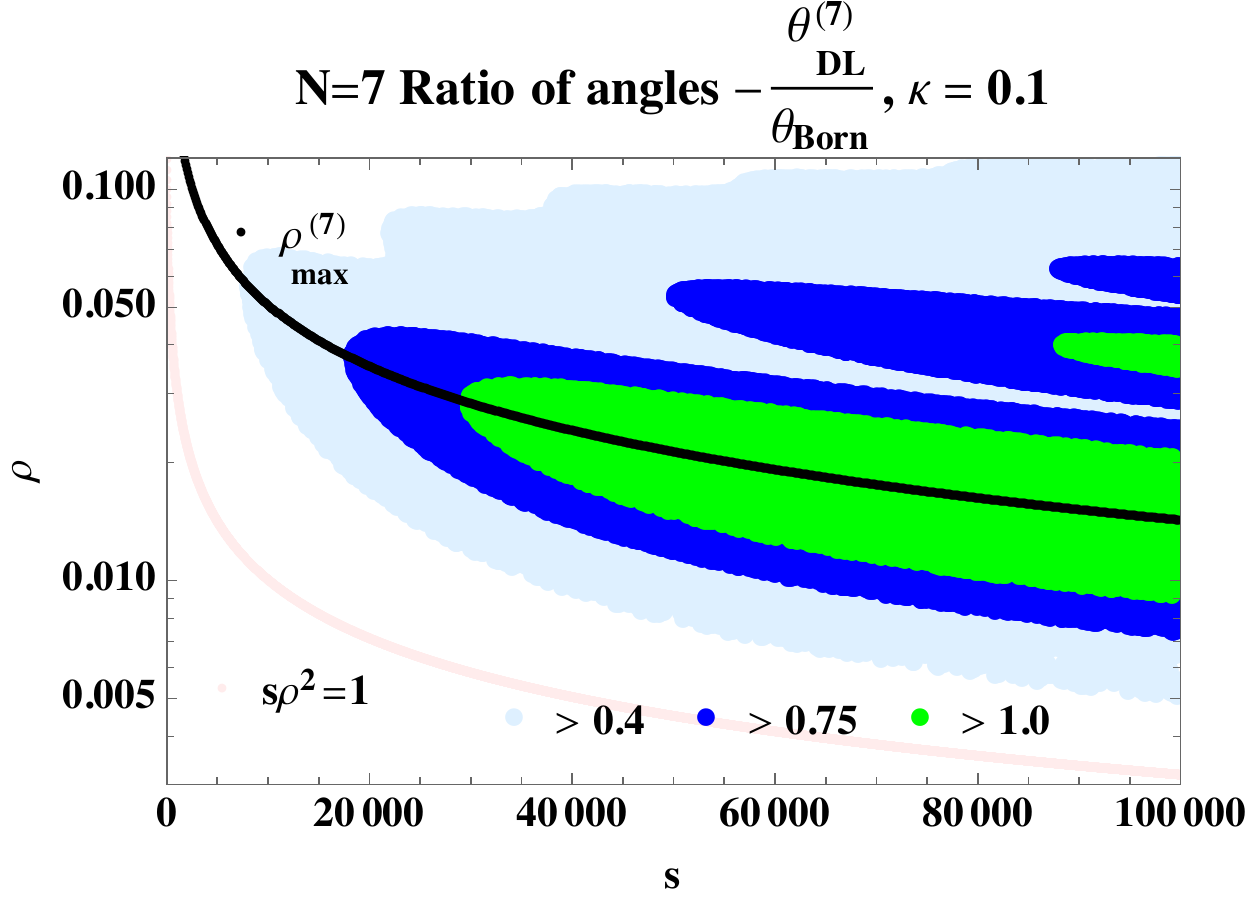}
\includegraphics[width=9cm]{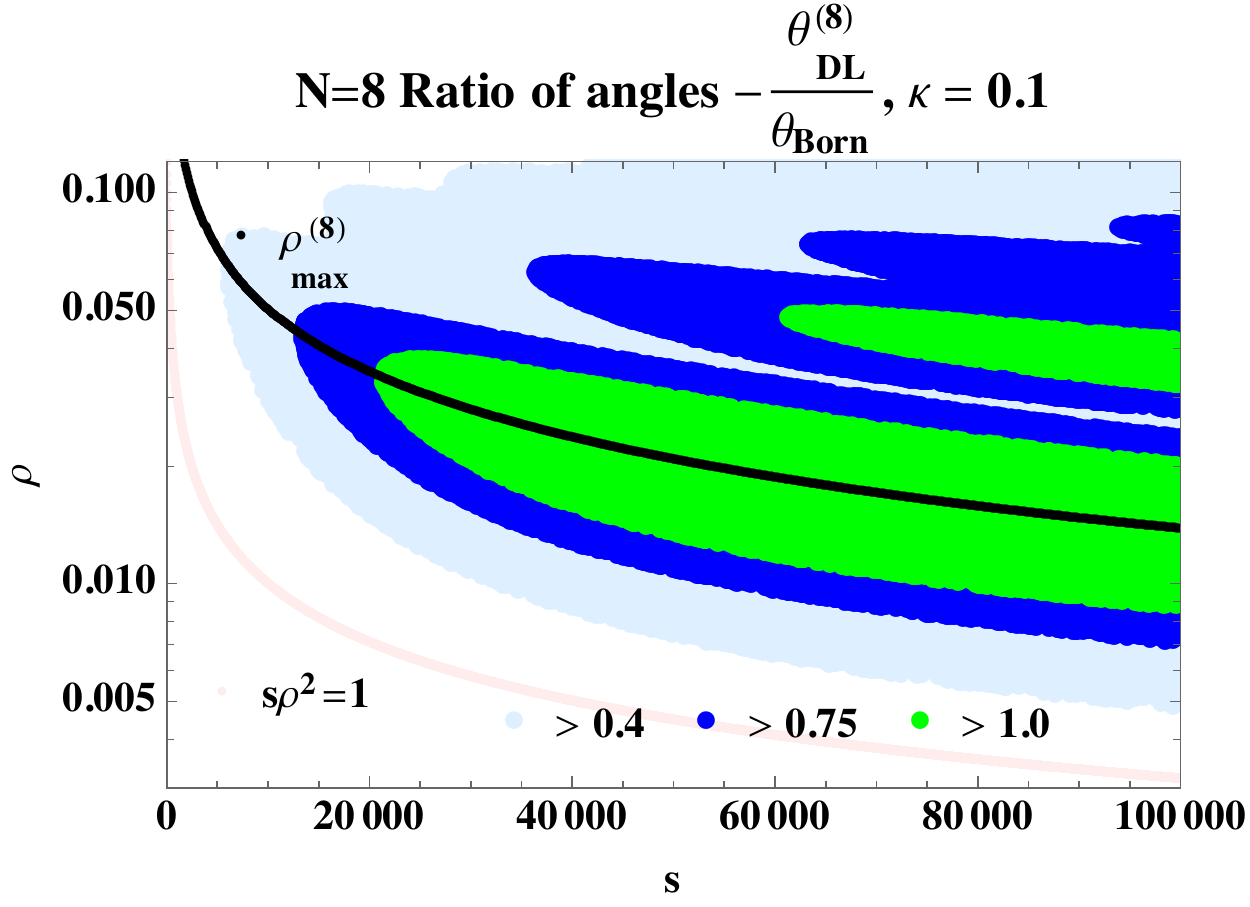}
\end{center}
\vspace{-.5cm}
\caption{Regions in impact parameter and energy space where the DL resummation effects are large ($N$=7 up, $N$=8 bottom). When the ratio $-\frac{\theta^{(N)}_{\rm DL}}{\theta_{\rm Born}}>1$ the deflection angle is negative. When it is bigger than 0.4 and 0.75, it corresponds to a large correction but still with a positive angle. }
\label{N78RatioAngles}
\end{figure}

The $s$ dependence of $\rho_{\rm max}^{(N)}$ is shown in Figs.~\ref{N56RatioAngles},~\ref{N78RatioAngles}, together with the ratio $-\frac{\theta^{(N)}_{\rm DL}}{\theta_{\rm Born}}$ 
in the regions close to $(s,\rho_{\rm max}^{(N)})$ and other areas in the $(s,\rho)$ plane where the ratio is large. Note that $\rho_{\rm max}^{(N)}$ is almost invariant under changes of $N$ for all ranges of energy. Three regions are highlighted corresponding to a negative correction of more than 40, 75 and 100 per cent of the Born value for the deflection angle. This shows how the largest 
corrections stemming from the DLs take place at small impact parameters and large energies. As already pointed out, they can even push the deflection angle to change sign when $-\frac{\theta^{(N)}_{\rm DL}}{\theta_{\rm Born}}>1$. These regions are larger in Fig.~\ref{N78RatioAngles}, for $N=7,8$, than in Fig.~\ref{N56RatioAngles}, with $N=5,6$. The line $s \rho^2 = 1$ is also included in the plots to show that the DL effects are relevant when $\rho > \frac{1}{\sqrt{s}}$, within the eikonal approximation of large $s$ and fixed impact parameter.  For $\rho > \rho_{\rm max}^{(N)}$ the deflection angle develops oscillations which manifest themselves as different bands in impact parameter space in Figs.~\ref{N56RatioAngles},~\ref{N78RatioAngles}. 

At this point it is important to remark  that this is a resummation of subleading terms. The purely eikonal contributions should indeed  play a major role in the  scattering process (see, {\it e.g},~\cite{Ciafaloni:2018uwe,DiVecchia:2019myk,DiVecchia:2019kta}). Already at one loop (see Eq.~(\ref{oneloopn8exact})) they are enhanced in the high energy limit since they carry a factor $\alpha \, s$ instead of $\alpha \, t$ for the DL terms here considered. Nevertheless the DL sector of the amplitude should not be neglected given that it strongly reduces the gravitational interaction at high energies and in a region of small impact parameters. The fact that this screening of gravity is more important for a larger number of gravitinos in the theory is presumably related to the possible lack of ultraviolet divergencies in ${\cal N}=8$  supergravity. It is noteworthy that this resummed sector of the four--graviton amplitude does not affect large distances, it is infrared finite, and thus it does not need of the regulator $\lambda$ present in the Born diagrams. 

The resummation of the DL terms therefore generates a $\lambda$--independent  negative shift for the eikonal phase in the short distance region which grows as the center--of--mass energy increases, it is absent for ${\cal N}=4$ supergravity  and maximal for ${\cal N}=8$ supergravity. This can be viewed as a negative contribution to the so-called Shapiro's time delay~\cite{Shapiro:1964uw}  experienced by a particle in a 
$2 \to 2$ scattering process~\cite{Kabat:1992tb}, which is akin to the time delay suffered by light travelling near a massive object when compared to its propagation in flat space.  The eikonal approximation in light--cone coordinates $(z^-,z^+,z^i)$, where the incoming particle $p_1$ ($p_2$) has 
a large $p^+$ ($p^-$) component, implies  
$s \simeq 2 p^+ p^-$,  $t \simeq - \vec{q}^{\, 2} = - q^2$.
The particle moving in the $z^+$ direction suffers a Shapiro's time delay which is related to the eikonal phase of Eq.~(\ref{phaseikonal}), {\it i.e.}
\begin{eqnarray}
\Delta (\rho, s) &=& \frac{\chi (\rho, s)}{|p^-|},
\end{eqnarray}
after its interaction with the other particle travelling in the $z^-$ direction. In ultraviolet complete theories it has been argued that, to avoid superluminality in the low energy limit, it is needed to allow only time delays and not time advances. This implies the positivity constraint $\chi (\rho, s) > 0$. Such a logic has been used to constrain the type of cubic vertices present in a new theory or to call for new physics at some intermediate scale to enforce the positivity condition on the eikonal phase~\cite{Camanho:2014apa,Hinterbichler:2017qyt,Bonifacio:2017nnt}. 

\begin{figure}[h]
\begin{center}
\includegraphics[width=9cm]{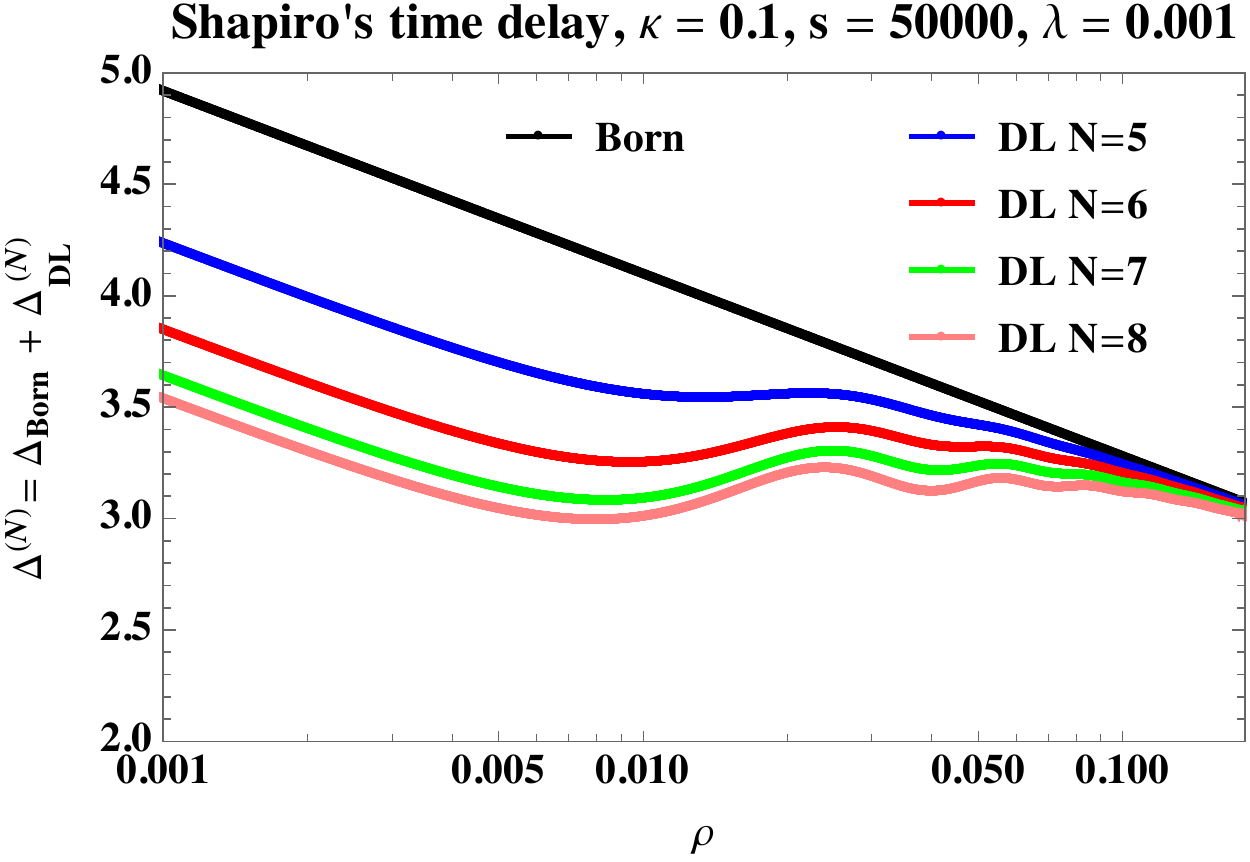}
\end{center}x
\vspace{-.5cm}
\caption{Shapiro's time delay for graviton--graviton scattering at Born level and for different supergravity theories with DL accuracy. The center of mass energy is chosen to be large, $s=50000$.}
\label{TimeDelaySeveralN}
\end{figure}

In the present context the Shapiro's time delay experienced by the gravitons at Born level would be
 \begin{eqnarray}
\Delta_{\rm Born} \left(\rho,s\right) &=&  
\frac{\partial }{\partial \sqrt{s}}  \chi_{\rm Born} \left(\rho,s\right)  ~=~ 
- \frac{ \kappa^2 \, \sqrt{s}}{2 \pi} \ln{\left(\rho \, \lambda \right)} \, .
\end{eqnarray}
The corresponding formula with DL corrections reads 
 \begin{eqnarray}
\Delta^{(N)} \left(\rho,s\right) &=&  \frac{\partial }{\partial \sqrt{s}}  \chi^{(N)} \left(\rho,s\right) ~=~   \Delta_{\rm Born} (\rho,s)  + 
 \Delta_{\rm DL}^{(N)} (\rho,s)  \nonumber\\
 &=&     \Delta_{\rm Born} (\rho,s)  +      2  \pi \alpha \sqrt{s} 
\sum_{m=1}^\infty    \sum_{n=0}^{m-1} 
\frac{  (m+1) (-\alpha s)^{m} {\cal C}_{m-n}^{(N)}}{(n!)^2 m^{2 (m-n) +1}}   
\left({\rho^2 \over 4 \alpha}\right)^{n}  \, .
\label{Timedelayone}
\end{eqnarray} 
A numerical study of this time delay is given in Fig.~\ref{TimeDelaySeveralN}. Very similarly as for the eikonal phase, the resummation of DL terms generates a constant shift for the function at very small impact parameters. In this region the shift of Shapiro's time delay becomes
 \begin{eqnarray}
\Delta^{(N)} \left(\rho,s\right) &\overset{ \rho \ll 1}{\simeq}&  
 \Delta_{\rm Born} (\rho,s)  
+      2  \pi \alpha \sqrt{s} 
\sum_{m=1}^\infty   
\frac{  (m+1) (-\alpha s)^{m} {\cal C}_{m}^{(N)}}{ m^{2 m +1}}  \\
&=&  \frac{\kappa^2\, \sqrt{s}}{4 \pi}\left(
  -  2\ln{\left(\rho \, \lambda \right)}   
+       \int_0^\infty\mu^{(N)} (x) \, dx
\sum_{m=1}^\infty     \left(1+\frac{1}{m}\right) \left(\frac{-\alpha s x}{m^2}\right)^{m}  \right)\, .  \nonumber
\label{}
\end{eqnarray} 
All quantum corrections contribute to this function at high energies to generate a result which is always positive in the $(s,\rho)$ regions explored in this work (Figs.~\ref{N56RatioAngles},~\ref{N78RatioAngles}). This implies that it is possible to have a screening of the gravitational interaction or even a repulsive gravitational potential without violating  causality in the DL eikonal limit. 

\section{Conclusions}

The double logarithmic in the center--of--mass energy, $s$, contributions to the four--graviton scattering amplitude have been investigated for four--dimensional 
${\cal N} \geq 4$ supergravities.

A novel integral representation for the coefficients of the perturbative expansion has 
been obtained in Eqs.~(\ref{coeffsCnN}, \ref{muN5}, \ref{muN6}, \ref{muN7}, \ref{muN8}). These should be compared to the current and future calculations of the higher order terms in the amplitude. Already the results here presented are  in agreement at two loops for ${\cal N}=4,5,6,8$ supergravities with those of~\cite{BoucherVeronneau:2011qv} and 
at three loops in ${\cal N}=8$  supergravity with~\cite{Henn:2019rgj}. The study of the associated singularities for the $t$--channel partial wave in the complex angular momentum plane has been reviewed and the asymptotic representation of the amplitudes  using the rightmost poles has been calculated with the correct normalization factors (Eqs.~(\ref{M4N5}, \ref{M4N6}, \ref{M4N7}, \ref{M4N8})).  

Following and improving the work in~\cite{SabioVera:2019edr}, the amplitude has been studied in impact parameter representation. The corrections to the eikonal phase have been written in terms of the perturbative coefficients in Eq.~(\ref{chidelcoeffs}). For small impact parameters they generate a negative constant  shift in the phase which is sensitive to a higher number of  DL  quantum corrections as the center of mass energy grows. 

Taking the derivative w.r.t. the impact parameter of the eikonal phase it is possible to study the graviton's deflection angle in the forward limit. The DL corrections to it, written in Eq.~(\ref{ThetaDLN}), are always negative and grow with energy. For some regions of small impact parameters and very large energies they can generate a negative deflection angle. This can be interpreted as a strong contribution to a screening of the gravitational interaction between both gravitons due to the presence of gravitinos in the quantum corrections to the scattering process. 

The derivative of the eikonal phase w.r.t. the scattering energy generates the Shapiro's time delay for each of the gravitons interacting at very high velocity. 
It has been shown that for small impact parameters this time delay receives a negative constant shift also sensitive to all DL quantum corrections. The explicit formula is given in Eq.~(\ref{Timedelayone}). The above mentioned regions with 
negative deflection angle present positivity for both the total eikonal phase and 
Shapiro's time delay and therefore do not have problems related to the violation of causality.

In~\cite{Kabat:1992tb} it was shown how Shapiro's time delay can be calculated in the context of a particle propagating in a shockwave background generated by another very fast moving particle. The Born level time delay was then obtained from an Aichelburg--Sexl metric~\cite{Aichelburg:1970dh,Dray:1984ha}. It would be interesting to search for the metric associated to the DL resummation here discussed. A further path for research is to calculate the DL terms of inelastic amplitudes, and also for ${\cal N}<4$  supergravities, where the DL amplitudes have a divergent behaviour for large $s$. The mapping to the problem of counting 1--rooted ribbon graphs found in~\cite{SabioVera:2019edr} for ${\cal N}=8$ supergravity should be understood in detail and, if possible, extended to other theories. Work is in progress to investigate these points.

\section*{Acknowledgements}

This work has been supported by the Spanish Research Agency (Agencia Estatal de Investigaci\'on) through the grant IFT Centro de Excelencia Severo Ochoa SEV-2016-0597, and the Spanish Government grants FPA2015-65480-P, FPA2016-78022-P.

\end{document}